\documentclass{article}
\usepackage{epsfig}
\usepackage[centertags]{amsmath}
\usepackage{amssymb}
\usepackage{graphicx}
\usepackage[square,numbers,sort&compress]{natbib}

\begin{document}

\title{Avoided Crossings in Driven Systems}

\author{ Benjamin P. Holder and Linda E. Reichl\\
Center for Studies in Statistical Mechanics and Complex Systems,\\
The University of Texas at Austin, Austin, Texas 78712}

\date{August 8, 2005}

\maketitle

\begin{abstract}
We characterize the avoided crossings in a two-parameter, time-periodic
system which has been the basis for a wide variety of experiments.  By
studying these avoided crossings in the near-integrable regime, we are able
to determine scaling laws for the dependence of their characteristic
features on the non-integrability parameter.  As an application of these
results, the influence of avoided crossings on dynamical tunneling is
described and applied to the recent realization of multiple-state tunneling
in an experimental system.

\end{abstract}


%
%
\section{\label{sec:1} Introduction}

Avoided crossings of eigenvalue curves are generic features of quantum
systems with non-integrable classical counterparts \cite{vonNeumannWigner}.
Their appearance allows for a wide variety of interesting, purely quantum
mechanical, phenomena including chaos-assisted tunneling \cite{tomsovic},
the adiabatic exchange of eigenstate character \cite{na_reichl}, and
generally provides the mechanism by which underlying classical chaos affects
the dynamics of a quantum system \cite{timberlake}.  Their existence is also
responsible for perhaps the most well-known result in the field of quantum
chaos, the non-Poisson statistical distribution of level spacings in
``chaotic'' quantum systems \cite{reichl_book}.

In systems with two parameters, an avoided crossing along any curve in
parameter space can be associated to a ``diabolical point'' at which two
eigenvalue {\em surfaces} become degenerate \cite{diabolical}.  The conical
shape of the two eigenvalue surfaces in the vicinity of such a diabolical point
ensures the characteristic hyperbolic behavior of two eigenvalues along any
curve in parameter space passing near, but not through, the diabolical
point.  In the particular case of a near-integrable system, one parameter
may be fixed to be zero leaving the system integrable for all values of the
other parameter.  Eigencurves will freely cross under variation of the latter
parameter, thus creating diabolical points of the associated eigenvalue
surfaces when viewed in the full two-parameter space.  As we show here, this
type of diabolical point is important because perturbation theory can be
applied to characterize the conical shape and therefore characterize the
avoided crossings of the near-integrable system.

In this paper we study the particular two-parameter, near-integrable
system of a harmonically driven pendulum:
\begin{equation}
H(\kappa,\lambda) = p^2 + \kappa \cos \theta + \lambda \left[ \cos
  \left(\theta + \omega t \right) + \cos \left( \theta - \omega t \right)
  \right] \,.
\end{equation}
This ``one-and-a-half'' degree-of-freedom system is one of the simplest
types of classical systems to exhibit chaos.  It is of significant
experimental interest in quantum mechanics since it has been implemented in
a number of studies
\cite{gs_zoller,raizen_dyn_loc,raizen_science,nist1,steck_raizen} through
the use of cold atom optics, particularly in investigations of
multiple-state dynamical tunneling \cite{steck_raizen}. Theoretically, it
provides a convenient framework for studying the avoided crossings of
near-integrable systems since for $\lambda \rightarrow 0$ the system is the
integrable pendulum Hamiltonian.

In the following, we study the properties of avoided crossings for the
driven pendulum with the use of Floquet theory.  An avoided crossing
of two Floquet eigenvalue curves (for $\lambda \ne 0$) can be
associated to a level crossing of the integrable pendulum
($\lambda=0$) system and is characterized by the dependence of its
closest approach on the non-integrability parameter $\lambda$. For
small values of $\lambda$, we find that the spacing exhibits a power
law dependence with an integer power.  A modified degenerate
perturbation theory is then applied to verify this dependence and
associate it to the direct or indirect coupling of the associated
integrable eigenstates.  We then use the perturbation results to
elucidate a multiple state dynamical tunneling process in the vicinity
of an avoided crossing and apply the results to the particular
achievement of this tunneling in an atom optics experiment.  We
finally show the association of this avoided crossing to a nearby
diabolical point.

In Section \ref{sec:model_ham} we present the model Hamiltonian under
consideration in the paper, including a description of the system's
classical dynamics.  Section \ref{sec:quant_dyn} presents the quantum
dynamics of the model system, with a brief review of Floquet analysis.
Avoided crossings of the model system are investigated in detail in Section
\ref{sec:AC}, first numerically, then with the perturbation theory results
presented in Appendix \ref{sec:pert_th}.  We review the implications of
avoided crossings on dynamical tunneling in Section \ref{sec:dyn_tunnel} and
demonstrate the origin of those avoided crossings in a particular
experimental system.  Section \ref{sec:conclusions} contains some concluding
remarks.

\section{\label{sec:model_ham}The Model Hamiltonian}

The Hamiltonian we consider consists of a particle moving in the
presence of a harmonically-modulated, spatially-periodic
potential.  It can be written in the form
\begin{equation}
H'(p',x,t') =  \frac{{p^{\prime}}^2}{2m} + V_1\cos(kx)+ V_2 \cos(kx)\cos({\omega}'t'),
\label{ham1}
\end{equation}
where $p'$ is the momentum and $x$ the position of a particle of mass $m$,
$t'$ is time, $V_1$ is the amplitude of the spatially periodic potential,
$V_2$ is the amplitude of the modulation potential and ${\omega}'$ is the
frequency of the modulation potential. The experimental implementation of
quantum systems of with this type of Hamiltonian was first proposed by Graham,
Schlautmann, and Zoller in 1992 \cite{gs_zoller} and then achieved by Raizen
{\em et. al.} \cite{raizen_dyn_loc,raizen_science,steck_raizen}, and Hensinger
{\em et. al.} \cite{nist1}.

It is useful to change to dimensionless units. We define:
$p=\frac{p^{\prime}}{\hbar k}$, $\theta=kx$,
$t=t^{\prime}\frac{E_0}{\hbar}$, $\omega=\omega^{\prime} \frac{\hbar}{E_0}$,
$\kappa= \frac{V_1}{E_0}$, $\lambda=\frac{V_2}{2 E_0}$, and
$H=\frac{H^{\prime}}{E_0}$, where $E_0\equiv \frac{\hbar^2 k^2}{2m}$.  Then,
the Hamiltonian in Eq. \ref{ham1} takes the form
\begin{equation}
H(p,\theta,t) = H_0(p,\theta) + \lambda \left[ \cos(\theta-{\omega}t)+
\cos(\theta+{\omega}t) \right],
\label{eqn:model-1}
\end{equation}
where
\begin{equation}
H_0(p,\theta) = p^2 + \kappa \cos(\theta),
\label{eqn:model-2}
\end{equation}
is the Hamiltonian of a pendulum and we have written the modulation term
explicitly as two travelling waves. Note that momentum is measured in units
of ${\hbar}k$.

\begin{figure}[!h]
     \begin{center}
     \includegraphics{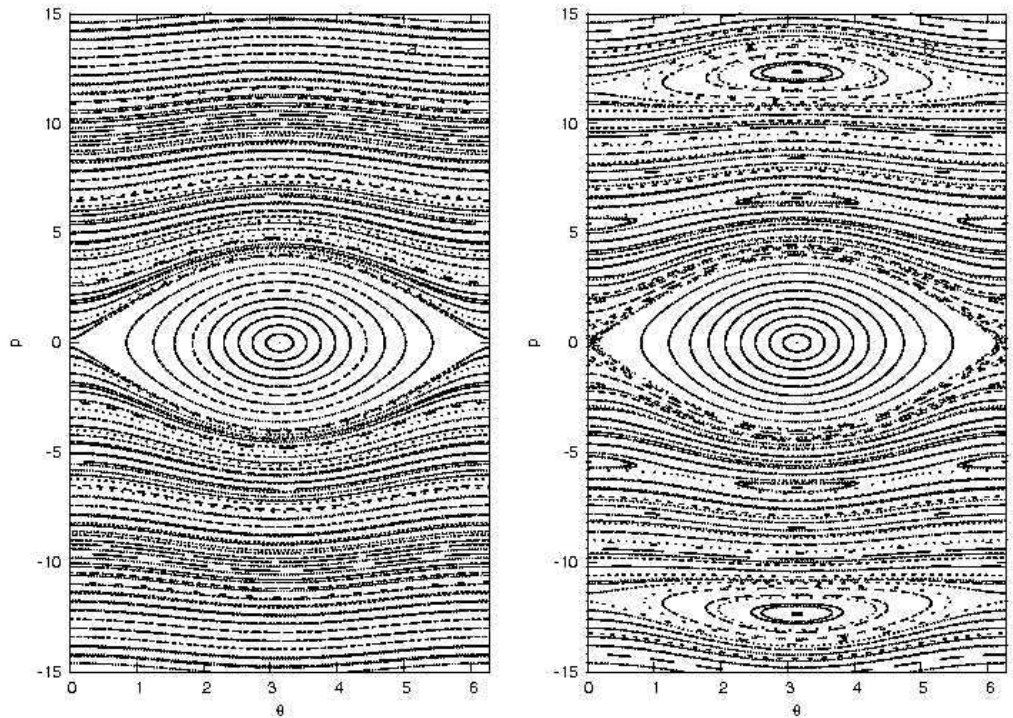}
     \end{center}
     \caption{Strobe plots of the system in Eq. (\ref{eqn:model-1}) with
$\omega=24$ and parameters: (a) $\kappa = 7.8$ and
$\lambda=0$; and (b) $\kappa= 7.8$ and $\lambda=1.0$.}
\label{fig:strobe}
\end{figure}

The classical phase space of a time-periodic one-and-a-half
degree-of-freedom system such as $H(p,\theta,t)$ can be visualized by a
strobe plot of the trajectories at times $t=\nu \frac{2 \pi}{\omega}
\;\; (\nu \in Z)$.  A strobe plot of phase space trajectories for the
system governed by Hamiltonian $H(p,\theta,t)$ is shown in Figure
\ref{fig:strobe}.a with parameters $\kappa=7.8$ and $\lambda=0$. (For
the case of $\lambda=0$, the system is independent of time and could
be visualized with an ordinary parametric plot of phase space, however
we plot the strobed phase space for convenience of comparison to the
perturbed system).  Because this system is integrable, all orbits lie on
tori (in either the regions of the pendulum's libration or rotation)
and the phase space is absent of chaos.  Figure \ref{fig:strobe}.b
shows a strobe plot of the phase space with parameters $\kappa=7.8$,
$\lambda=0.5$, and $\omega=24$.  The travelling waves in the
modulation term have phase velocities $v=\pm \omega$ and are seen as
{\em primary resonance} structures at $p=\pm \frac{\omega}{2}$ where $
\dot \theta = v$.  Although much of the orbit structure of the
integrable system is preserved, the tori with rational winding numbers
have been destroyed, giving rise to a self-similar set of {\em
daughter resonance} structures (see, for example, the two-island chains
at $p=\pm \frac{\omega}{4}$). Regions of chaos surround these
resonances, most visibly near the separatrix of the pendulum
resonance.

\section{\label{sec:quant_dyn}Quantum Dynamics}

The dimensionless Schr\"odinger equation for the system in
Eq. (\ref{eqn:model-1}) is $i \frac{\partial}{\partial t}
|{\psi}(t)\rangle = \hat{H}(t) |{\psi}(t)\rangle$, where
\begin{equation}
\label{eqn:quant_model_scale}
     \quad \quad \hat{H}(t) = \hat{p}^2 + \kappa \cos\hat{\theta} + 2
\lambda \cos\left(\hat{\theta}\right) \, \cos(\omega t) \,,
\end{equation}
We will consider the configuration space, $\theta \in [0,2 \pi)$, to
be periodic such that $\langle \theta + 2 \pi|\psi(t)\rangle = \langle
\theta |\psi(t) \rangle$ and the momentum operator has integer
eigenvalues: $\hat p |p\rangle = n |p \rangle \;\; (n \in Z)$.  In the
experimental systems, this is approximately achieved naturally because
momentum transfer occurs in discrete units of $\hbar k$
\cite{raizen_science,nist1,luter_reichl}.

When $\lambda=0$, the Hamiltonian reduces to that of the quantum pendulum,
$\hat{H}_0(\kappa) = \hat{p}^2 + \kappa \cos\hat{\theta}$.  The
eigenstates of $\hat{H}_0$ are Mathieu functions \cite{abramowitz} which we
will henceforth denote as $|n (\kappa) \rangle$ so that $\hat{H}_0(\kappa)
|n (\kappa)\rangle = E_n(\kappa) |n (\kappa)\rangle$ where $n = 0, \pm 1,\pm
2, \ldots$.  (We will suppress the $\kappa$-dependence of these eigenstates
until their specification is necessary).  States $|n\rangle$ with positive
integer labels are those with even parity, states with negative integer
label are those with odd parity. Note that as $\kappa \rightarrow 0$, $E_n
\rightarrow n^2$.  If $\kappa \ne 0$, but $|n|$ is large (i.e. the
corresponding classical pendulum energy is much larger than that of the
separatrix), we will again have $E_n \approx n^2$.

\subsection{Floquet Theory }
\label{sec:floq_states}

The Hamiltonian in Eq. (\ref{eqn:quant_model_scale}) is time-periodic
and therefore Floquet's theorem guarantees that solutions of the
Schr\"odinger equation can be written in the form
\begin{equation}
|\psi_{\alpha}(t)\rangle={\rm e}^{- i\Omega_{\alpha}
 t}|\phi_{\alpha}(t)\rangle \quad {\rm with} \quad | \phi_{\alpha}(t + T) \rangle = |
 \phi_{\alpha} (t) \rangle \,,
 \label{eqn:floq_state}
\end{equation}
where we have defined $T=\frac{2 \pi}{\omega}$ and where
$|\phi_{\alpha}(t) \rangle$ and $\Omega_{\alpha}$ are called the
Floquet eigenstate and eigenvalue, respectively. Substituting this
solution into the Schr\"odinger equation yields the eigenvalue
equation
\begin{equation}
\label{eqn:floq_ham}
\hat{H}_F(t) |\phi_{\alpha}(t) \rangle \equiv \left( \hat{H}(t) - i
\frac{\partial}{\partial t} \right) |\phi_{\alpha} (t) \rangle =
\Omega_{\alpha} | \phi_{\alpha}(t) \rangle\,,
\end{equation}
where $\hat{H}_F(t)$ is called the Floquet Hamiltonian.

The Floquet Hamiltonian is a Hermitian operator in a {\em composite Hilbert
space} $\Theta \otimes \mathcal{T}$ \cite{sambe,dzyublik}, where
$\Theta$ is the space of all square-integrable functions $f(\theta)$ on the
configuration space and $\mathcal{T}$ is the space of all time-periodic
functions $a(t)$ with period $T$ and finite $\int_{-T/2}^{T/2} |a(t)|^2 dt$.
The inner product of two vectors $|\phi_a\rangle$ and $|\phi_b\rangle$ in
this space is then defined by
\begin{equation}
\langle\langle \phi_a | \phi_b \rangle \rangle \equiv \frac{1}{T}
\int_{-T/2}^{T/2} \langle \phi_a| t \rangle \langle t | \phi_b
\rangle dt = \frac{1}{T} \int_{-T/2}^{T/2} \langle \phi_a(t) |
\phi_b(t) \rangle dt\,,
\end{equation}
where $\langle \phi_a(t) | \phi_b(t) \rangle = \int_{0}^{2 \pi}
\langle \phi_a(t) | \theta \rangle \langle \theta | \phi_b(t) \rangle
d\theta$ is the usual inner product in $\Theta$.  We select
a complete orthonormal basis in this composite space
\begin{equation}
\langle t | n,q\rangle = | n \rangle \; {\rm e}^{i q \omega t} \quad \quad \left( n,q \in
 Z \right)\,,
\end{equation}
where $\left\{ |n \rangle \right\}$ are the eigenstates of the
pendulum Hamiltonian $\hat{H}_0$. These basis vectors satisfy $\langle
\langle n,q | n^{\prime}, q^{\prime} \rangle \rangle =
\delta_{n,n^{\prime}} \delta_{q,q^{\prime}}$.

The Floquet Hamiltonian $\hat{H}_F$ is Hermitian, so the Floquet eigenvalues
$\Omega_{\alpha}$ are real and two Floquet eigenstates $|\phi_{\alpha}
\rangle$ and $| \phi_{\beta} \rangle$ belonging to different eigenvalues are
orthogonal. Additionally, the Floquet Hamiltonian commutes with the parity
operator defined by its action on the momentum eigenket $\hat{\Pi} | p
\rangle = |-p \rangle$.  Therefore the two operators can be diagonalized
simultaneously and all Floquet eigenstates have definite parity: $\hat{\Pi}
| \phi_{\alpha} \rangle = \pm 1 | \phi_{\alpha} \rangle $. Floquet states
with parity eigenvalue $+1$ will be called even, those with eigenvalue $-1$
odd.

Given one Floquet eigenstate $| \phi_{\alpha}(t) \rangle $ with Floquet eigenvalue
$\Omega_{\alpha}$, there will be another Floquet eigenstate $|
\phi^{\prime}_{\alpha} (t)\rangle$ such that $| \phi^{\prime}_{\alpha}(t)
\rangle \equiv {\rm e}^{i q \omega t} | \phi_{\alpha}(t) \rangle \;\; (q
\in Z)$, with eigenvalue $\Omega_{\alpha}^{\prime} \equiv \Omega_{\alpha} + q
\omega$. These two Floquet eigenstates, however, represent the same physical
state, i.e.
\begin{equation}
{\rm e}^{-i \Omega^{\prime}_{\alpha} t} | \phi^{\prime}_{\alpha}(t)
\rangle = {\rm e}^{-i \Omega_{\alpha} t} | \phi_{\alpha}(t) \rangle \,.
\end{equation}
Therefore we may limit consideration to the {\em fundamental zone} $- \omega/2
\le \Omega < \omega/2$ in which each physical eigenstate of the time-dependent
Schr\"odinger equation is represented by the corresponding Floquet eigenstate
with eigenvalue within that range.

\begin{figure}
\begin{center}
\includegraphics{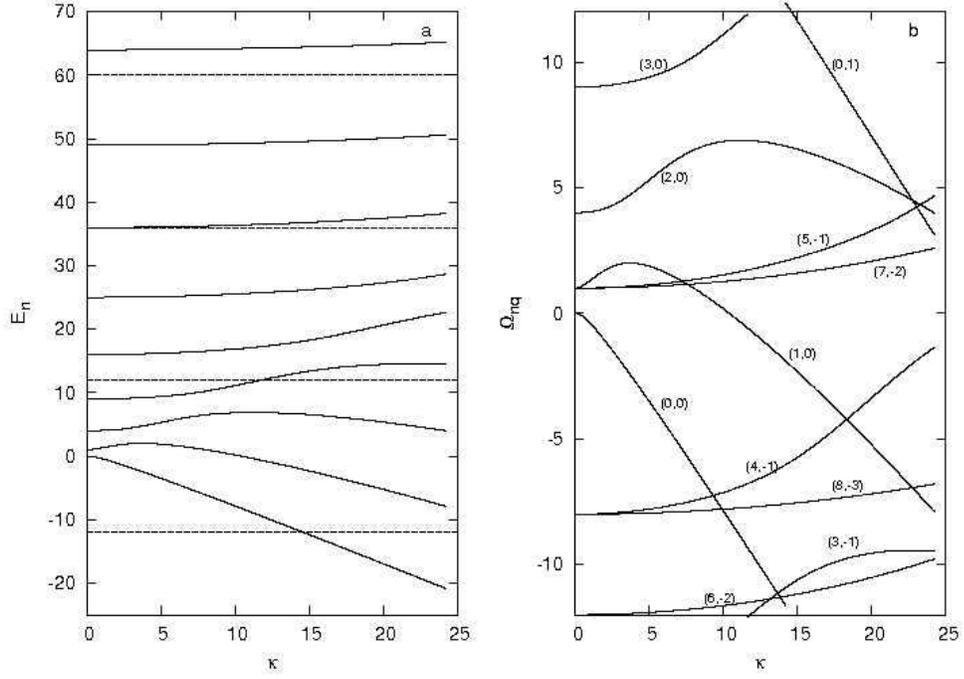}
\end{center}
\caption{(a) Energy curves of the nine lowest-energy, even-parity
eigenstates of the quantum pendulum, $\hat{H}_0 = \hat{p}^2 + \kappa \cos
\hat{\theta}$. (b) The nine corresponding ``Floquet pendulum'' eigenvalues
in the fundamental zone $-\frac{\omega}{2} < {\Omega} \leq +
\frac{\omega}{2} $ with $\omega=24$. The labels $(n,q)$ on each Floquet
eigenvalue segment identify the corresponding Floquet eigenstate $|n,q
\rangle$.  The dashed lines in (a) indicate the Floquet eigenvalue zone
boundaries.} \label{fig:pendulum_to_floquetpend}
\end{figure}

Consider the unperturbed system $\hat{H}_F^0 \equiv \hat{H}_0 - i
\frac{\partial}{\partial t}$ which we will call the {\em Floquet
pendulum}.  The eigenstates of this system are precisely the basis
states $|n,q \rangle$ with eigenvalues $\Omega_{nq} = E_n + q \omega$.
Figure \ref{fig:pendulum_to_floquetpend} shows the lowest nine energies
of the even-parity eigenstates of the quantum pendulum and the
corresponding Floquet eigenvalues in the fundamental
zone $-\omega/2 \le \Omega < \omega /2$.

\subsection{\label{sec:floq_mat}Another method for determining Floquet states}

An arbitrary dynamical state of the system can be expanded, with the use of
equation (\ref{eqn:floq_state}), in the basis of Floquet eigenstates,
\begin{equation}
|\psi(t) \rangle = {\sum_{\alpha}}^{\prime} A_{\alpha} {\rm e}^{-i
\Omega_{\alpha} t} | \phi_{\alpha} (t) \rangle \,,
\end{equation}
where the ``prime'' indicates that the sum is restricted to those Floquet
states with $\Omega_{\alpha}$ in the fundamental zone.  The expansion
coefficients are independent of time and can be written $A_{\alpha} =
\langle \phi_{\alpha}(0) | \psi (0) \rangle$. Using the time-periodicity of
the Floquet eigenstates, we can then write
\begin{equation}
| \psi(T) \rangle = \sum_{\alpha}^{\prime} {\rm e}^{-i \Omega_{\alpha} T} |
\phi_{\alpha}(0) \rangle \langle \phi_{\alpha}(0) | \psi(0) \rangle \equiv
\hat{\rm U}(T) | \psi(0) \rangle \,,
\end{equation}
showing that the time-evolution operator over a single period $T$
\begin{equation}
\hat{\rm U}(T) = {\sum_{\alpha}}^{\prime} {\rm e}^{-i \Omega_{\alpha} T} |
\phi_{\alpha}(0) \rangle \langle \phi_{\alpha}(0) |
\end{equation}
is diagonalized by the Floquet eigenstates at time $t=0$.  We can therefore
determine these {\em time-strobed Floquet states} by constructing the matrix
${\rm U}_{mm^{\prime}} \equiv \langle m | \hat{\rm U}(T) | m^{\prime}
\rangle $ in some convenient basis $\{| m \rangle \}$ in $\Theta$,
truncating this matrix at some appropriate level $m=M$ where it becomes
approximately diagonal (i.e. ${\rm U}_{MM} >> {\rm U}_{Mm}$ for $ m \ne M$),
and then performing a numerical diagonalization to obtain the
$|\phi_{\alpha} (0) \rangle$ and $\Omega_{\alpha}$ (mod $\omega$).  The
$m^{th}$ column of ${\rm U}$ is obtained by evolving the basis vector
$|m\rangle$ over one period $T$ via numerical integration of the
Schr\"odinger equation.

In subsequent sections, we will compare the phase space distributions of the
time-strobed Floquet eigenstates $| \phi_{\alpha} (0) \rangle$ to the classical
system. We can do this by introducing the Husimi distribution $\rho
(\theta_0,p_0)$ \cite{husimi,knaufsinai} of a quantum state
$| \phi \rangle$ on the classical phase space $(\theta_0,p_0)$
\begin{equation}
\rho(\theta_0,p_0) \equiv \frac{1}{2 \pi} | \langle \theta_0,p_0 | \phi
\rangle |^2 \,,
\end{equation}
where the {\em coherent state} $|\theta_0,p_0 \rangle$ is defined as an
eigenstate of the annihilation operator $\hat a = \frac{1}{\sqrt{2}} \left(
\hat{\theta}/{\sigma} + i \sigma \hat{p} \right)$ with position and momentum
expectation values of $\theta_0$ and $p_0$ respectively.  The free parameter
$\sigma$ is set according to the physical system considered (see below).
The representation of such a coherent state in the discrete momentum basis
$\{ | p \rangle \}$ is given by
\begin{equation}
\langle p | \theta_0, p_0 \rangle = A \exp \left[ - \frac{\sigma^2}{2}
(p-p_0)^2 - i \theta_0 (p - p_0) \right]\,,
\end{equation}
where $A$ is a normalization factor guaranteeing $\langle \theta_0,p_0 |
\theta_0,p_0 \rangle = 1$.  The action of the annihilation operator on the
coherent state can be used to show that $\langle p \rangle \equiv \langle
\theta_0,p_0|\hat{p}|\theta_0,p_0 \rangle = p_0$, $\langle \theta \rangle =
\theta_0$, $\Delta \theta = \sigma/\sqrt{2}$, and $\Delta p = (\sigma
\sqrt{2})^{-1}$. Thus, the coherent state is a minimum-uncertainty wavepacket,
where the free parameter determines the ratio of its uncertainty in position
and momentum, i.e. $\sigma^2 = \Delta \theta / \Delta p$.  Reference
\cite{knaufsinai} presents an in-depth discussion on the selection of the
parameter $\sigma$. In all Husimi plots shown in subsequent sections, we set
$\sigma = 1.18 \, \kappa^{-1/4}$, a choice which provides the best association
between the quantum pendulum eigenfunctions and the corresponding classical
orbits).  As we will see, the Husimi distributions of the Floquet states lie
directly on the orbit structures of the classical phase space.

\section{\label{sec:AC}Avoided crossings}

The Floquet pendulum is integrable and its eigenvalues
$\Omega_{n,q}(\kappa)$, shown in Figure \ref{fig:pendulum_to_floquetpend}.b,
cross under the variation of $\kappa$.  For any nonzero $\lambda$, however,
the system represented by the Hamiltonian in Eq.
(\ref{eqn:quant_model_scale}) is non-integrable and the approach of any two
(same parity) Floquet eigenvalues under variation of $\kappa$ results in an
avoided crossing.  This well-known result, the {\em no-crossing theorem},
was first proven by von Neumann and Wigner for eigenvalues of generic
Hermitian matrices \cite{vonNeumannWigner}.  They also showed that adiabatic
passage of two quantum states through an avoided crossing leads to an
exchange of character. (In a two-parameter system, this exchange can be
related to the partial circuit of a diabolical point \cite{diabolical},
while in single parameter systems it can be related to exceptional points in
the complex parameter plane \cite{heiss}). Avoided crossings of Floquet
eigenvalues in the fundamental zone, which generally involve states
localized in well-separated regions of the phase space, will therefore allow
a wide variety of interesting quantum dynamical phenonomena, including
adiabatic transitions and tunneling.

In this section, we will consider the near-integrable regime ($0 < \lambda <<
\kappa$), in which a clear association can be made between the Floquet
eigenstates of the perturbed system ($\lambda \ne 0$) and those of the Floquet
pendulum ($\lambda=0$). In this regime, the Floquet eigenvalues will follow
nearly the same dependence on $\kappa$ as the unperturbed eigenvalues seen in
Figure \ref{fig:pendulum_to_floquetpend}.b, except in the vicinity of an
avoided crossing. For $\kappa$ values sufficiently far from these avoided
crossings, we can make a unique, though necessarily local, association $|
\phi_{\alpha} \rangle \leftrightarrow | n_{\alpha},q_{\alpha} \rangle$ of the
Floquet eigenstate $|\phi_{\alpha} \rangle$ to the Floquet pendulum state with
maximum overlap $|\langle n,q | \phi_{\alpha} \rangle |$.  As $\lambda
\rightarrow 0$, this association will become an equality. We will see that the
fundamental characteristics of an avoided crossing between states
$|\phi_{\alpha} \rangle$ and $| \phi_{\beta} \rangle$ will be determined by the
difference
\begin{equation}
\Delta q_{\alpha \beta} \equiv | q_{\alpha} - q_{\beta} |\,.
\end{equation}
In the subsections below, we first present a numerical analysis of some
representative avoided crossings in the system with $\omega=24$, and then use
perturbation theory to show that the results are quite general.

\subsection{\label{sec:ACnum}Numerical Results}
\begin{figure}
     \begin{center}
     \includegraphics{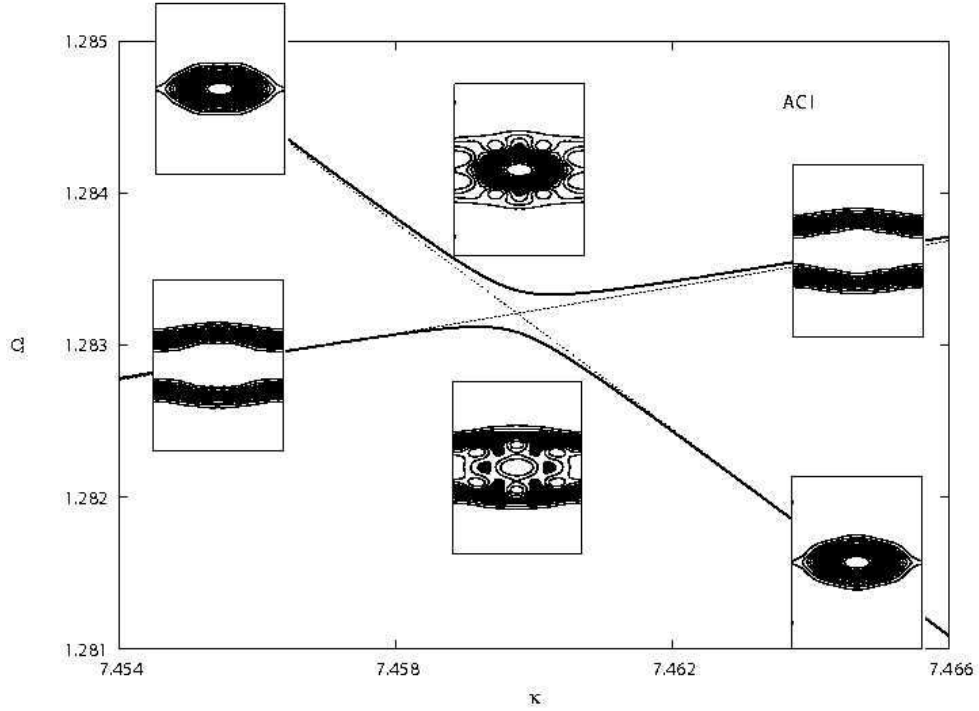}
     \end{center}
     \caption{A $\Delta q_{\alpha \beta}=1$ avoided crossing of the system
     $\hat{H}_F$ with parameters $\omega=24$ and $\lambda=5 \times 10^{-2}$.
     The Husimi distributions of the corresponding Floquet eigenstates are
     shown at $\kappa = (7.455,7.460,7.465)$ (the horizontal axis is
     $\theta_0 \in [0,2 \pi)$, vertical is $p_0 \in [-15,15]$).  The dotted
     lines are the eigencurves of the unperturbed Floquet pendulum.}
     \label{fig:ACI}
\end{figure}
\begin{figure}
     \begin{center}
     \includegraphics{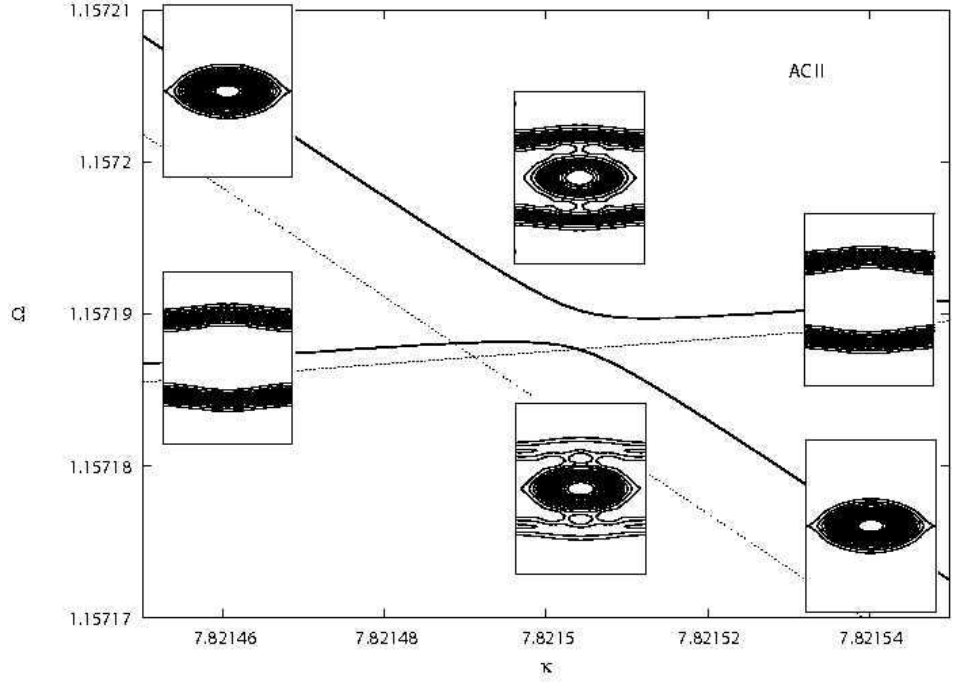}
     \end{center}
     \caption{A $\Delta q_{\alpha \beta}=2$ avoided crossing of the system
$\hat{H}_F$ with the same $\lambda$ and $\omega$ values as in Figure
\ref{fig:ACI}. The Husimi distributions shown are the two Floquet eigenstates
at $\kappa = (7.82146,7.82150,7.82154)$ (axes are the same as Figure
\ref{fig:ACI}).  The dotted lines are the eigencurves of the unperturbed
Floquet pendulum.} \label{fig:ACII}
\end{figure}
\begin{figure}
     \begin{center}
     \includegraphics{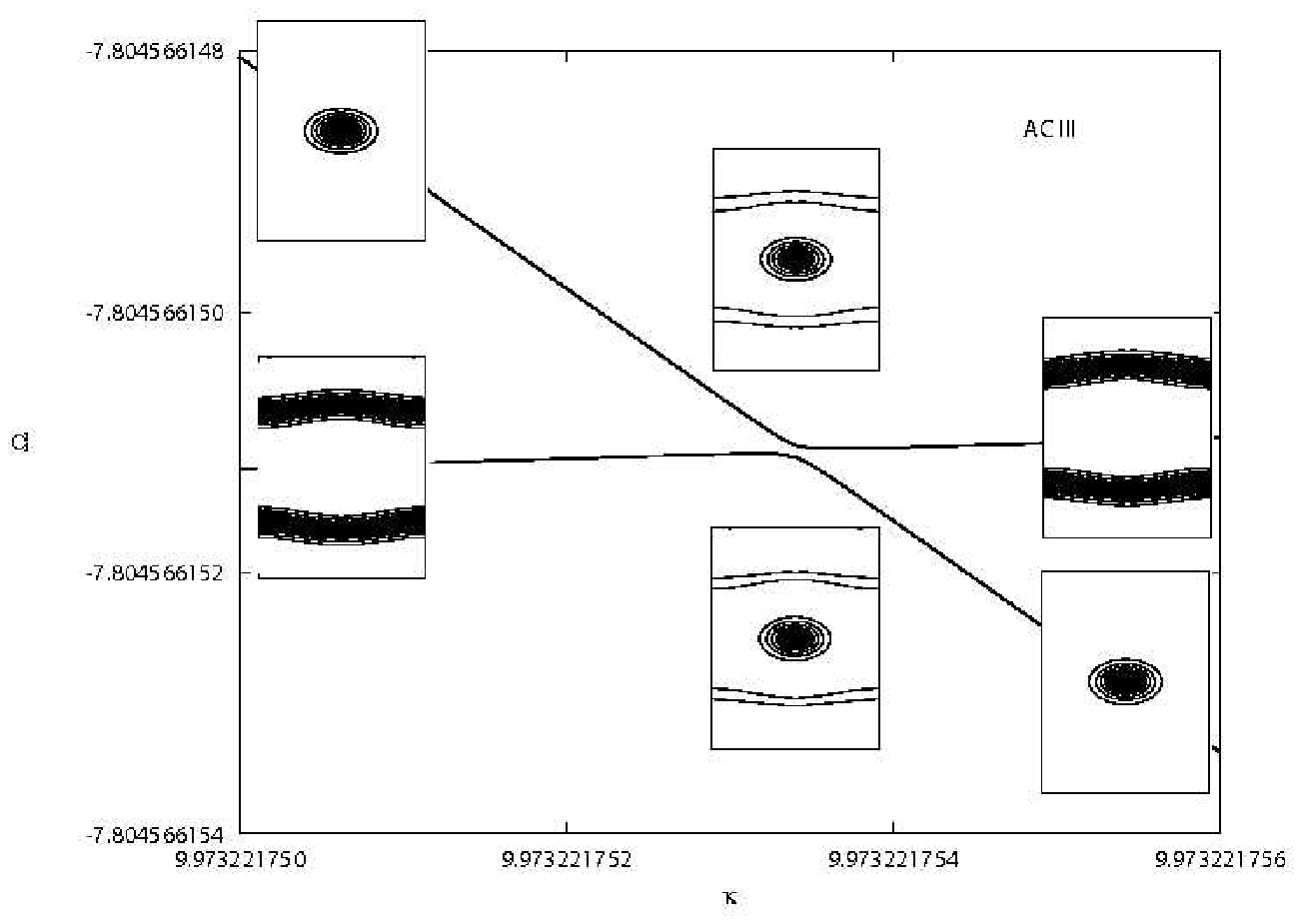}
     \end{center}
     \caption{A $\Delta q_{\alpha \beta}=3$ avoided crossing of the system
     $\hat{H}_F$ with the same $\lambda$ and $\omega$ values as in Figure
     \ref{fig:ACI}. The Husimi distributions shown are the two Floquet
     eigenstates at $\kappa = (9.973221751,9.9732217534,9.973221755)$ (axes
     are the same as Figure \ref{fig:ACI}).  The crossing of the
     corresponding eigencurves of the
     unperturbed Floquet pendulum falls outside of the plotted region
     at $\kappa_0 \approx 9.973242$.}
     \label{fig:ACIII}
\end{figure}

Three avoided crossings of Floquet eigenvalues in the fundamental zone
with $\lambda=5 \times 10^{-2}$ are shown in Figures \ref{fig:ACI},
\ref{fig:ACII}, and \ref{fig:ACIII} with Husimi plots of the
corresponding Floquet eigenstates overplotted on the figures. The
dotted lines shown are the eigenvalues of the unperturbed Floquet
pendulum. These plots were created by numerically calculating the
time-strobed Floquet states at a sequence of $\kappa$ values. With
each step forward in $\kappa$, the new states were associated to those
of the previous step by calculating the maximum overlap and verifying
continuity of the eigenvalues.  In the case that two Floquet
eigenvalues crossed between $\kappa$-steps, the size of the step was
reduced and the process repeated until no crossing occurred.

Each of these three avoided crossings involves one Floquet eigenstate localized
within the pendulum resonance at $p=0$ and another localized outside of the
pendulum resonance.  The avoided crossing in Figure \ref{fig:ACI} involves the
states $(n_{\alpha},q_{\alpha})=(1,0)$ and $(n_{\beta},q_{\beta}) = (5,-1)$;
Figure \ref{fig:ACII} involves the states $(n_{\alpha},q_{\alpha})=(1,0)$ and
$(n_{\beta},q_{\beta}) = (7,-2)$; and Figure \ref{fig:ACIII} involves states
$(n_{\alpha},q_{\alpha})=(0,0)$ and $(n_{\beta},q_{\beta}) = (8,-3)$.  The
associated crossings can be found in Figure
\ref{fig:pendulum_to_floquetpend}.b. These particular avoided crossings were
chosen as representative examples with $\Delta q_{\alpha \beta} = 1$, $2$, and
$3$, respectively.

Some general characteristics of these avoided crossings deserve attention.
First, the ``exchange of character'' between the two states is evident in the
evolution of the Husimi distributions with $\kappa$.  The associations
$|\phi_{\alpha} \rangle \leftrightarrow |n_{\alpha},q_{\alpha} \rangle$ and
$|\phi_{\beta} \rangle \leftrightarrow |n_{\beta},q_{\beta} \rangle$ well {\em
before} the avoided crossing become $|\phi_{\alpha} \rangle \leftrightarrow
|n_{\beta},q_{\beta} \rangle$ and $|\phi_{\beta} \rangle \leftrightarrow
|n_{\alpha},q_{\alpha} \rangle$ well {\em after}.  For $\kappa$ values at the
avoided crossing, the two Floquet states are superpositions of the asymptotic
states.  Second, there is quite a disparity of scale among the three avoided
crossings.  In particular, the minimum eigenvalue spacing $\Delta_{\alpha
\beta}$, defined by
\begin{equation}
\Delta_{\alpha \beta} \equiv  {\rm min} \left( |\Omega_{\alpha} -
\Omega_{\beta}|
\right)\,,
\end{equation}
is relatively large for the first and relatively small for the third.
Finally, the position $\kappa_{ac}$ of the minimum spacing (which we will
henceforth call the ``position of the avoided crossing''), is not
necessarily equal to the position $\kappa_0$ of the unperturbed crossing.
Indeed, it seems that for $\Delta q_{\alpha \beta} \ne 1$, the avoided
crossing is significantly offset in both $\kappa$ and $\Omega$.

\begin{figure}
     \begin{center}
     \includegraphics{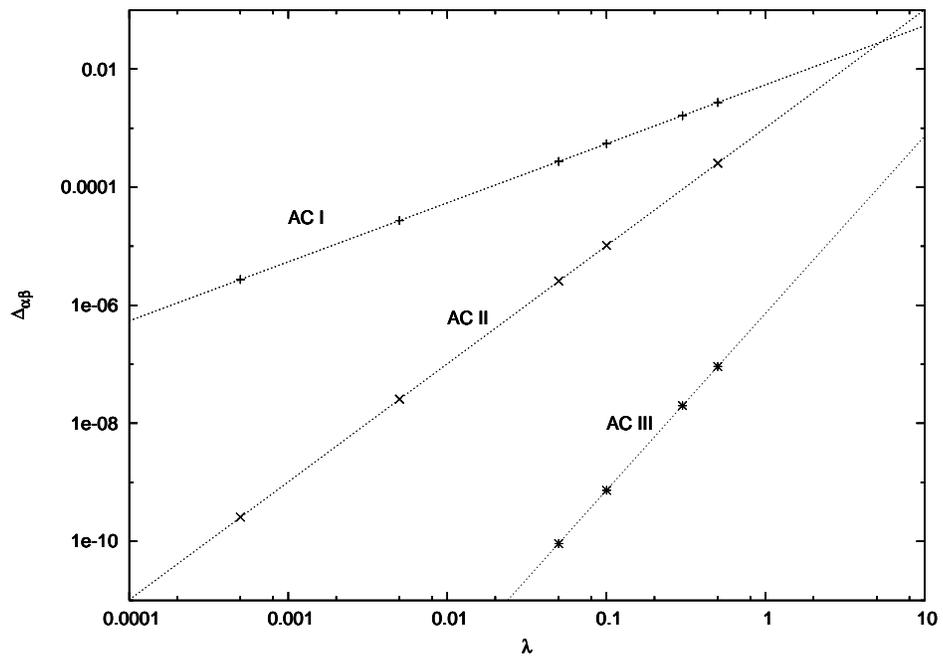}
     \end{center}
     \caption{The minimum spacing $\Delta_{\alpha \beta}$ as a function of
     $\lambda$ of the three avoided crossings shown in Figures
     \ref{fig:ACI}, \ref{fig:ACII}, and \ref{fig:ACIII}. The functions
     $\Delta_{\alpha \beta} = 5.44 \times 10^{-3} \lambda$, $\Delta_{\alpha
     \beta} = 1.02 \times 10^{-3} \lambda^2$, $\Delta_{\alpha \beta} = 7.35
     \times 10^{-7} \lambda^3$ are overplotted.} \label{fig:delta_vs_lambda}
\end{figure}

\begin{figure}
     \begin{center}
     \includegraphics{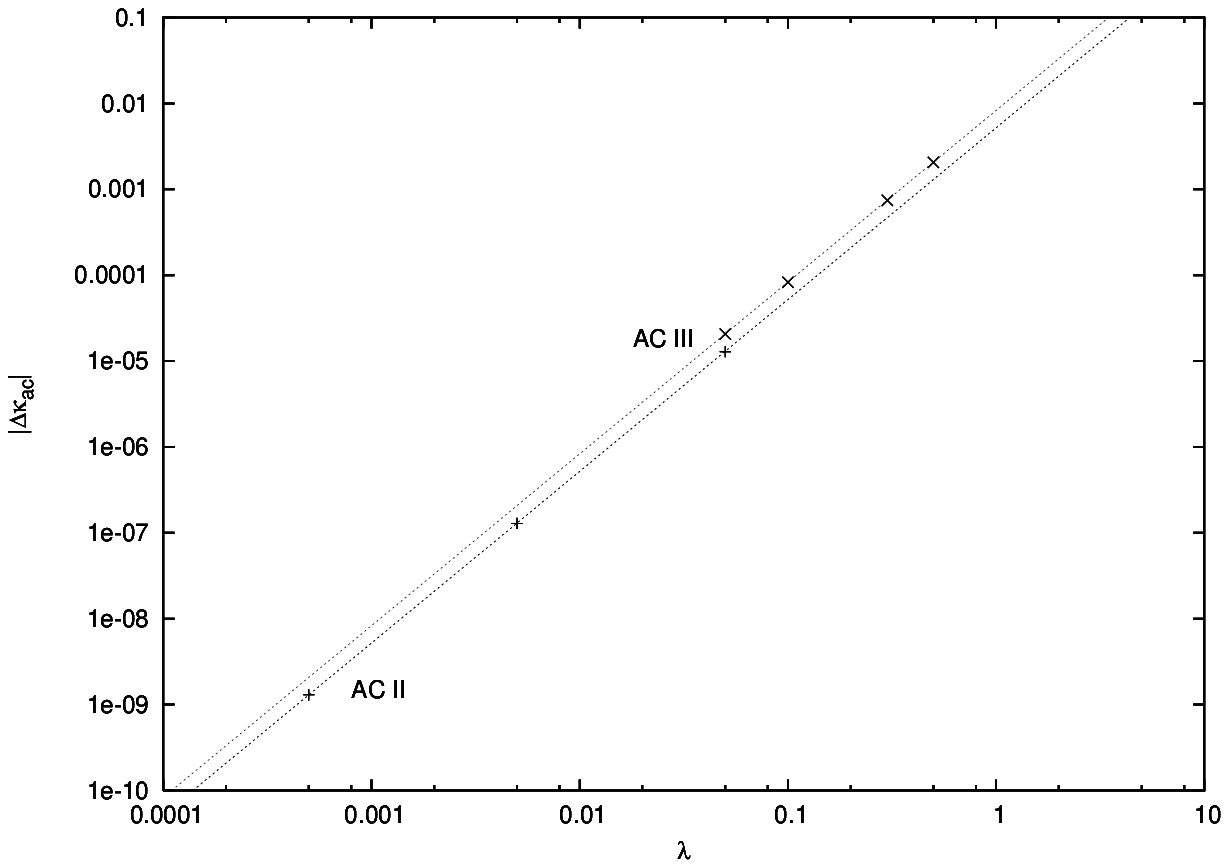}
     \end{center}
     \caption{The $\kappa$-offset $\Delta\kappa_{ac}$ as a function of
$\lambda$ for the avoided crossings shown in Figures \ref{fig:ACII}
and \ref{fig:ACIII}. The functions $|\Delta\kappa_{ac}| = 5.2 \times
10^{-3} \; \lambda^2$ and $|\Delta\kappa_{ac}| = 8.3 \times 10^{-3} \;
\lambda^2$ are overplotted.}
\label{fig:dkappa_vs_lambda}
\end{figure}

To make these last two observations more quantitative, we have computed
dependence of $\Delta_{\alpha \beta}$ and $\Delta\kappa_{ac} \equiv |\kappa_0 -
\kappa_{ac}|$ on the parameter $\lambda$. The results for the three example
avoided crossings are shown in Figures \ref{fig:delta_vs_lambda} and
\ref{fig:dkappa_vs_lambda}. We see that, for small values of $\lambda$, the
dependences are all well approximated by power laws with integer exponents. The
minimum spacing of the avoided crossings is given by
\begin{equation}
\label{eqn:Delta_vs_lambda}
\Delta_{\alpha \beta} = A \; \lambda^{\Delta q_{\alpha \beta}}\,,
\end{equation}
where the coefficients are $A \approx \{5.44 \times 10^{-3},1.02
\times 10^{-3},7.35\times 10^{-7}\}$ for avoided crossings I,II, and
III, respectively. The $\kappa$-offset of the avoided crossings are
given by
\begin{equation}
\label{eqn:Dkappa_vs_lambda}
\Delta\kappa_{ac} = B \; \lambda^2 \,,
\end{equation}
where $B \approx \{ 5.2 \times 10^{-3},-8.3 \times 10^{-3}\}$ for avoided
crossings II and III ($\Delta \kappa_{ac} = 0$ for avoided crossing I).
\subsection{\label{sec:pert_th_res}Perturbation theory results}

We now use perturbation theory to determine the behavior of the
Floquet eigenvalues and eigenstates in the neighborhood of avoided
crossings.  We will obtain approximate solutions ($|\phi_{\alpha}
\rangle,\Omega_{\alpha}$) to the Floquet eigenvalue equation
\begin{equation}
\label{eqn:floq_eig} \hat{H}_F(\kappa,\lambda) |
\phi_{\alpha}(\kappa,\lambda) \rangle = \left( \hat{H}^0_F(\kappa) + \lambda
\hat{V} \right) |\phi_{\alpha} (\kappa,\lambda) \rangle =
\Omega_{\alpha}(\kappa,\lambda) | \phi_{\alpha}(\kappa,\lambda) \rangle \,,
\end{equation}
where $\hat{H}^0_F(\kappa)$ is the Floquet pendulum Hamiltonian, $\hat{V} =
2 \cos\hat{\theta} \,\cos(\omega t) $, and $\lambda$ is considered a small
expansion parameter.  Our unperturbed system is the two-fold degenerate
system $\hat{H}^0_F(\kappa_0)$, where $\kappa_0$ is the parameter value at
which the eigenvalue curves of two Floquet pendulum states $|\alpha^0
(\kappa) \rangle \equiv |n_{\alpha},q_{\alpha} \rangle$ and $|\beta^0
(\kappa)\rangle \equiv |n_{\beta}, q_{\beta} \rangle$ cross.  We have seen
in Section \ref{sec:ACnum} that, for $\lambda \ne0$, the closest approach of
the eigenvalues $\Omega_{\alpha}$ and $\Omega_{\beta}$ involved in an
avoided crossing may not occur at $\kappa = \kappa_0$, so an offset must be
allowed for. We therefore introduce into (\ref{eqn:floq_eig}) an arbitrary
function $\kappa(\lambda)=\kappa_0 + \Delta\kappa(\lambda)$ and expand
$\Delta\kappa(\lambda)$ as a power series in $\lambda$.  The particular
value $\kappa_{ac}(\lambda)$ at which the eigenvalues make their closest
approach can then be determined by solving the extremal condition for
$\Delta \Omega_{\alpha \beta} \equiv | \Omega_{\alpha}(\kappa,\lambda) -
\Omega_{\beta}(\kappa,\lambda)|$,
\begin{equation}
\label{eqn:extremal}
\left. \frac{\partial \Delta \Omega_{\alpha \beta}}{\partial \kappa}
\right\rvert_{\kappa=\kappa_{ac}} = 0 \,,
\end{equation}
at each order to fix the expansion coefficients of $\Delta \kappa(\lambda)$. In
this manner, we find the perturbed eigenstates and eigenvalues at
$\kappa=\kappa_{ac}$.

The   details   of   the   perturbation   analysis   are  given  in  Appendix
\ref{sec:pert_th}.  The results may  be summarized  as  follows.   The breaking
of the degeneracy between states $|\alpha^0 \rangle$ and $|
\beta^0 \rangle$ occurs at the lowest order $\lambda^N$ for which the
coupling between these states, $v^{(N)}_{\alpha  \beta}$ (defined below), is
nonzero.  At this order, the two Floquet eigenvalues in the region  of the
avoided crossing are determined by the eigenproblem  of a $2 \times 2$ matrix
in the basis of the unperturbed states $|\alpha^0 (\kappa_0) \rangle$ and $|
\beta^0 (\kappa_0)\rangle$:
\begin{equation}\label{eqn:2by2}
     \left( \begin{array}{cc}
     \Delta\kappa^{(N)} {\delta}E_{\alpha} + v^{(N)}_{\alpha \alpha} &
     v^{(N)}_{\alpha \beta} \\
     v^{(N)}_{\beta \alpha} &
     \Delta\kappa^{(N)}{\delta}E_{\beta} + v^{(N)}_{\beta \beta}
     \end{array} \right)
     \left( \begin{array}{c}
     C^{\pm}_{\alpha} \\
     C^{\pm}_{\beta}  \\
     \end{array} \right) =  \Omega_{\pm}^{(N)}
     \left( \begin{array}{c}
     C^{\pm}_{\alpha} \\
     C^{\pm}_{\beta}  \\
     \end{array} \right)\,.
\end{equation}
The coefficients $C^{\pm}_{\alpha}$ and $C^{\pm}_{\beta}$ determine the
zeroth-order near-degenerate eigenstates in the region of the avoided crossing; $\Delta
\kappa^{(N)}$ and $\Omega_{\pm}^{(N)}$ are the coefficients of $\lambda^N$
in the expansions of the arbitrary $\kappa$-offset and the near-degenerate
Floquet eigenvalues $\Omega_{\pm}$, respectively; $\delta E_i=\langle
n_i(\kappa_0) | \cos{\theta} | n_i (\kappa_0)\rangle$ are the slopes of the
unperturbed eigencurves; and $v^{(N)}_{ij}$ depends on the matrix elements
of the perturbation operator:
\begin{equation}
V_{lm} \equiv \langle \langle n_l,q_l| \hat{V}| n_m,q_m \rangle \rangle =
\langle n_l(\kappa_0) | \cos \hat{\theta} | n_m(\kappa_0) \rangle \left(
\delta_{q_l,q_m+1} + \delta_{q_l,q_m - 1} \right) \,.
\end{equation}
For the first three orders in $\lambda$, these couplings are
\begin{gather}
v_{ij}^{(1)} = V_{ij} \\
v_{ij}^{(2)} = \sum_{\gamma \not\in \{ \alpha,\beta
\}}\frac{ V_{i \gamma} V_{\gamma j}}{\Omega^{(0)} - \Omega^{(0)}_{\gamma}}
\\
v_{ij}^{(3)} = \sum_{\gamma \not\in \{ \alpha,\beta \}} \left\{
\sum_{\sigma \not\in \{ \alpha,\beta \}} \frac{ V_{i \gamma} V_{\gamma
\sigma} V_{\sigma j}}{\left(\Omega^{(0)} -
\Omega^{(0)}_{\gamma}\right)\left( \Omega^{(0)} -
\Omega^{(0)}_{\sigma} \right)} \right. \nonumber \\ \left. +
\delta_{ij} \frac{V_{i \gamma} V_{\gamma i} \left[V_{\alpha \alpha}
(\delta E_{\alpha} - \delta E_{\gamma}) + V_{\beta \beta} \left(
\delta E_{\gamma} - \delta E_{\beta} \right) \right]}{\left(
\Omega^{(0)} - \Omega^{(0)}_{\gamma} \right)^2} \right\} \,,
\end{gather}
where $i,j \in \{\alpha, \beta \}$, the zeroth-order eigenvalues
$\Omega^{(0)}_l$ are taken at $\kappa_0$, and we write $\Omega^{(0)} =
\Omega_{\pm}^{(0)}$.  Using Eqs. (\ref{eqn:2by2}) and (\ref{eqn:extremal})
we find that the $N^{th}$-order corrections to the eigenvalues, at the
position of the avoided crossing, are given by
\begin{equation} \label{eqn:OmegaN}
\begin{split}
\Omega^{(N)}_{\pm}(\kappa_{ac}) &= \frac{1}{2} \left[ v_{\alpha \alpha}^{(N)} +
  v_{\beta \beta}^{(N)} + \Delta \kappa^{(N)}_{ac} \left( \delta E_{\alpha}
  + \delta E_{\beta} \right) \right] \\
&\quad \pm \frac{1}{2}\sqrt{\left[
  v_{\alpha \alpha}^{(N)} - v_{\beta \beta}^{(N)} + \Delta
  \kappa^{(N)}_{ac} \left( \delta E_{\alpha} - \delta E_{\beta} \right)
  \right]^2 + 4 | v_{\alpha \beta}^{(N)}|^2} \\
&=\frac{v^{(N)}_{\beta \beta} - v^{(N)}_{\alpha
    \alpha}}{\delta E_{\alpha} - \delta E_{\beta}} \pm |
    v^{(N)}_{\alpha \beta}|\,.
\end{split}
\end{equation}
At orders $0<M<N$, the two eigenvalues $\Omega_{\pm}$ are degenerate at an
offset from $(\kappa_0, \Omega^{(0)})$ specified by the coefficients
\begin{equation}
\Delta \kappa^{(M)} = \frac{v^{(M)}_{\beta \beta} - v^{(M)}_{\alpha
    \alpha}}{\delta E_{\alpha} - \delta E_{\beta}}\,,
\end{equation}
and
\begin{equation}
\Omega^{(M)} = \frac{v^{(M)}_{\beta \beta} \delta E_{\alpha} - v^{(M)}_{\alpha
    \alpha} \delta E_{\beta}}{\delta E_{\alpha} - \delta E_{\beta}}\,.
\end{equation}

The origin of the numerical results presented in section \ref{sec:ACnum} is
now clear.  The matrix elements of the perturbation $V_{ij}$ are non-zero
only when $\Delta q_{ij}=1$.  Therefore, for an avoided crossing between
states $|\phi_{\alpha} \rangle$ and $|\phi_{\beta} \rangle$, we must have
$\Delta q_{\alpha \beta}=N$.  Using Eq. (\ref{eqn:OmegaN}) and the fact that
$\Omega^{(M<N)}$ is the same for $|\phi_{\alpha} \rangle$ and $|\phi_{\beta}
\rangle$, we find that, to lowest order in $\lambda$, the minimum spacing is
given by
\begin{equation}
\Delta_{\alpha \beta} = 2 \; \left\lvert \, v^{(\Delta q_{\alpha \beta})}_{\alpha
  \beta} \right\rvert \; \lambda^{\Delta q_{\alpha \beta}}\,.
\end{equation}
Its interesting to note that the $\kappa$-offset of an avoided crossing in
this system is dependent on $\lambda^2$, because $v^{(2)}_{\alpha \alpha}$
and $v^{(2)}_{\beta \beta}$ are non-zero, even when $\Delta q_{\alpha \beta}
> 2$.

\begin{table}
\begin{equation}
\begin{array}{|l|l|l|l|l|l|l|l|}
\hline \kappa_{ac} & (n_{\alpha},q_{\alpha}) & (n_{\beta},q_{\beta}) & \Delta
q_{\alpha \beta} & A_{\rm num} & A_{\rm pt} & B_{\rm num} &  B_{\rm pt} \\
\hline
7.46 & (1,0) & (5,-1) & 1 & 5.44 \times 10^{-3} & 5.5 \times 10^{-3} & ~~~--- &
~~~--- \\
7.83 & (1,0) & (7,-2) & 2 & 1.02 \times 10^{-3} & 1.0 \times 10^{-3} &
5.2 \times 10^{-3} & 5.2 \times 10^{-3} \\
9.97 & (0,0) & (8,-3) & 3 & 7.35 \times 10^{-7} &
7.3 \times 10^{-7} & -8.3 \times 10^{-3} & -8.3 \times 10^{-3} \\
\hline
\end{array} \nonumber
\end{equation}
\caption{Comparison of numerical results to those of perturbation
  theory.  The quantities $A_{\rm num}$ and $B_{\rm num}$ are obtained
  from numerical simulation and $A_{\rm pt}$ and $B_{\rm pt}$ are obtained
  from perturbation theory.}
\label{table:results}
\end{table}

Table \ref{table:results} shows a quantitative comparison of the numerical
results presented in the previous section (in terms of the coefficients $A$ and
$B$ of Eqs. (\ref{eqn:Delta_vs_lambda}) and (\ref{eqn:Dkappa_vs_lambda})) to
those obtained by the perturbation analysis. For all three examples of avoided
crossings, we see excellent agreement.  We have also verified the predictions
of perturbation theory for a number of other avoided crossings in this system
(as well as those in systems with different values of $\omega$), finding
similar agreement.

A number of other characteristics of the avoided crossings can be determined
from our perturbation analysis.  Substituting
$\Omega^{(N)}_{\pm}(\kappa_{ac})$ and $\Delta \kappa^{(N)}_{ac}$ into Eq.
(\ref{eqn:2by2}), we find that at the position of the avoided crossing, the
two perturbed Floquet eigenstates (to lowest order in $\lambda$) become an
equal superposition of the two associated Floquet pendulum states, i.e.
\begin{equation}
|C^{\pm}_{\alpha}(\Delta \kappa_{ac})| = |C^{\pm}_{\beta}(\Delta
\kappa_{ac})|\,.
\end{equation}
We may also determine the relative magnitudes of these coefficients at some
$\kappa$ value near the avoided crossing.  If, instead of calculating $\Delta
\kappa^{(N)}$ by the extremal condition, we instead determine the
$\kappa$-offset where the eigenvalue separation is $a$ times the minimum
value, we find 
\begin{equation}
\left.\Delta \kappa^{(N)}
\right\rvert_{\Delta\Omega_{\alpha \beta} = a \Delta_{\alpha \beta}} =
\frac{v_{\alpha \alpha}^{(N)} - v_{\beta \beta}^{(N)}}{\delta
  E_{\alpha \alpha} - \delta E_{\beta \beta}} \pm \frac{2 \sqrt{a^2 -
    1} \, | v^{(N)}_{\alpha \beta} |}{|\delta E_{\alpha \alpha}- \delta
  E_{\beta \beta}|} \,.
\end{equation}
A simple calculation then shows that the coefficients obey
\begin{equation}
\left[\frac{|C^{\pm}_{\alpha}|}{|C^{\pm}_{\beta}|}\right]_{\Delta\Omega_{\alpha
\beta}= a \Delta_{\alpha \beta}} = \frac{1}{|a \pm \sqrt{a^2 - 1}|}
\,,
\end{equation}
where we have assumed that $(\delta E_{\alpha} - \delta E_{\beta})>0$. As $a
\rightarrow (0,\infty)$, we see that, for example,
$\frac{|C^+_{\alpha}|}{|C^+_{\beta}|} \rightarrow (1,0)$, as expected. This
verifies the qualitative behavior seen in the Husimi distributions plotted
in Figures \ref{fig:ACI}-\ref{fig:ACIII}.

\section{\label{sec:dyn_tunnel}Implications for Dynamical Tunnelling}
Under the evolution of the system given in Eq.
(\ref{eqn:quant_model_scale}), an initial quantum state $|\psi_+(0)\rangle$
which is localized in classical phase space (in the sense of its Husimi
distribution) in a region of positive momentum at $p \approx p_0$ may
undergo time-periodic dynamical tunneling, across the central resonance (and
all intervening KAM tori) to the opposite momentum region at $p \approx
-p_0$. The mechanism for this behavior is the existence of a near-degenerate
and opposite parity pair of Floquet eigenstates which each have localization
near $p_0$ and $-p_0$. If the Floquet eigenvalues of these two states are
far from any avoided crossings, the tunneling dynamics are well described by
a two-state process exactly analogous to the tunneling through a potential
barrier in the time-independent double well system \cite{sakurai}.  In the
vicinity of an avoided crossing, however, the dynamics are influenced by a
third state with partial localization in the regions of $\pm p_0$ and the
time-evolution takes on a more complicated beating behavior.  In this
section we analyze this tunneling behavior in the perturbative regime
($\lambda$ small) and then apply the results to tunneling oscillations
observed in a recent experiment.  Although the experimental system cannot be
considered to be in a perturbative regime, we identify the diabolical point
associated to the relevant avoided crossing and show that an approximate
result can be obtained numerically which characterizes this avoided crossing
quite well.

\subsection{Tunneling behavior in the perturbative limit}
\begin{figure}
     \begin{center}
     \includegraphics{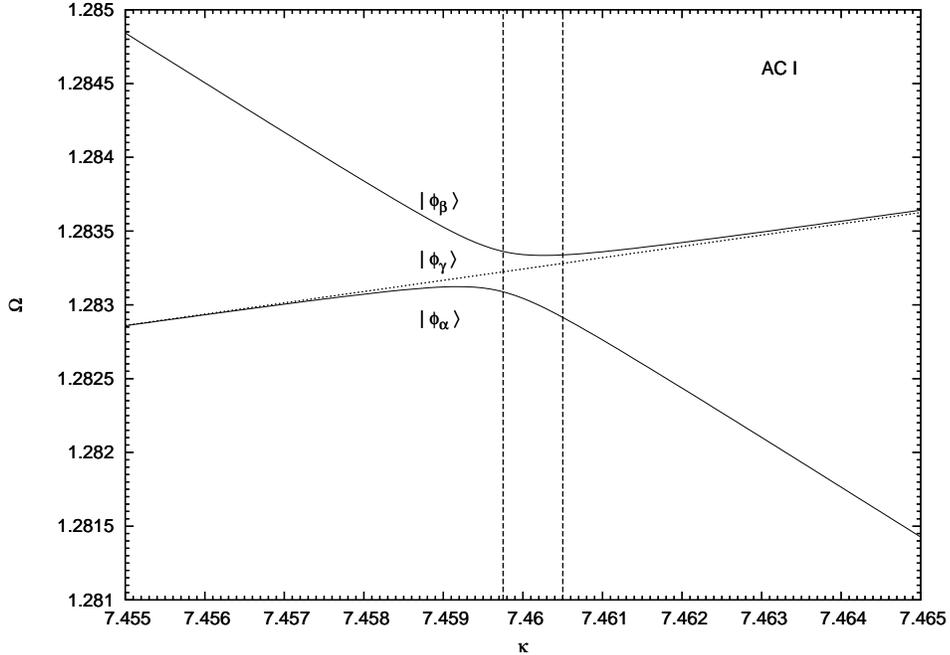}
     \end{center}
     \caption{The avoided crossing from Figure \ref{fig:ACI} (a), now with
the relevant odd-parity state included (dotted line).  The vertical lines
are the $\kappa$ values at which the time-evolution in Figure
\ref{fig:ACI_tunnel_evolve} was obtained.} \label{fig:ACI_tunnel}
\end{figure}

To analyze the tunneling induced by the Hamiltonian in Eq.
(\ref{eqn:quant_model_scale}), we consider again the one-period time-evolution
operator $\hat{\rm U}(T)$ defined in Section \ref{sec:floq_mat} and its
eigenvectors, the time-strobed Floquet eigenstates $| \phi_{\alpha}(T)
\rangle = | \phi_{\alpha}(0) \rangle$ (we will drop the explicit reference
to ``time-strobed'' in this section).  As a particular example, consider the
avoided crossing shown in Figure \ref{fig:ACI_tunnel} (the same as that
shown in Figure \ref{fig:ACI}, but with the odd-parity state now included,
shown as a dotted line).  At $\kappa=7.456$, a value far from the avoided
crossing, the opposite-parity pair of states $|\phi_{\alpha} \rangle$ and $|
\phi_{\gamma} \rangle$ are near-degenerate ($|\Omega_{\gamma} -
\Omega_{\alpha}| \approx 4.5 \times 10^{-6}$) and have localization at $p
\approx \pm 5$. The phase space localization of state $|\phi_{\beta}
\rangle$ is completely within the central resonance and does not overlap
significantly with this pair.  We can construct an initial state, localized
at either $p \approx 5$ or $-5$, as an equal superposition of the two
near-degenerate Floquet states, i.e. $|\psi_{\pm}(0) \rangle =
\frac{1}{\sqrt{2}} \left( | \phi_{\alpha} \rangle \pm | \phi_{\gamma}
\rangle \right)$.  When either is acted on by $\hat{\rm U}(nT)$, the
evolution is periodic, oscillating between $p \approx \pm 5$ with a
tunneling frequency $\omega_{tun} = | \Omega_{\alpha} - \Omega_{\gamma}|$.
The corresponding number of modulation periods for complete oscillation is
then $n_{tun}=\omega / \omega_{tun} \approx 5.3 \times 10^6$.

\begin{figure}
     \begin{center}
     \includegraphics{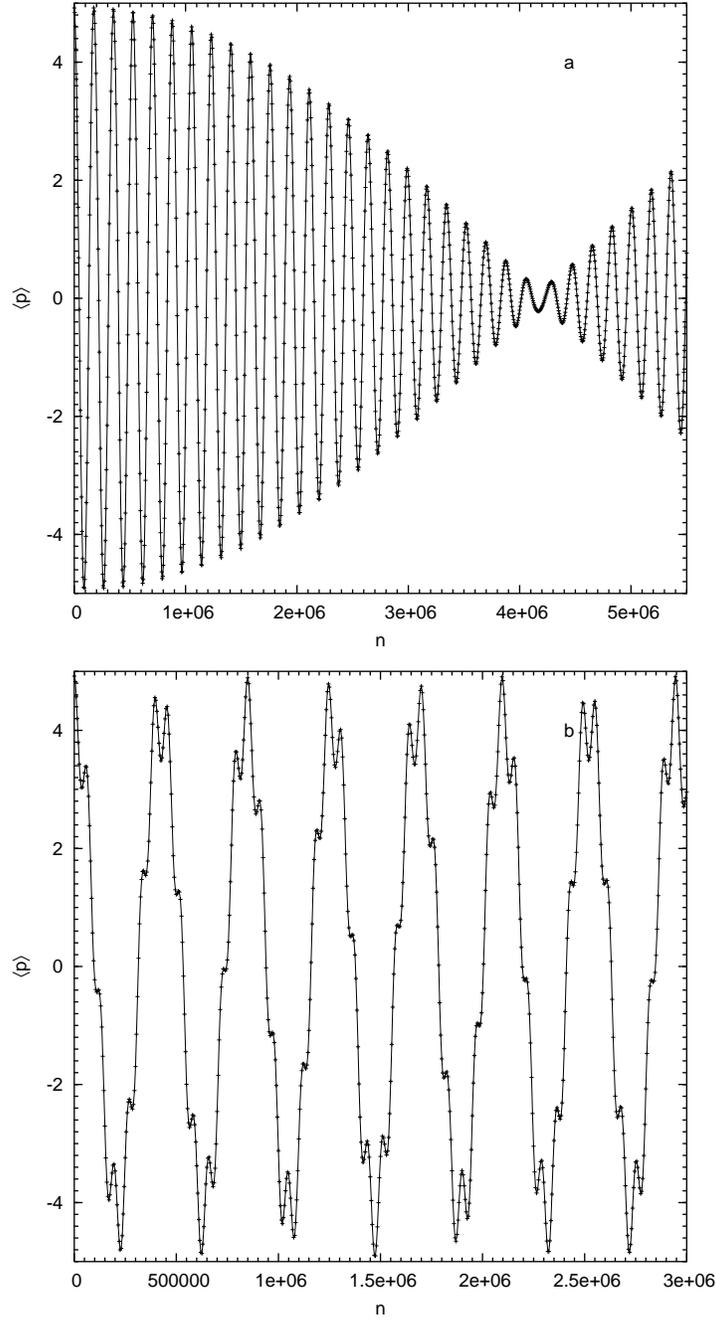}
     \end{center}
     \caption{The evolution of state $| \psi_+(0) \rangle$ (created by the
     superposition of the near-degenerate states at $\kappa=7.456$) under
     the action of $\hat{\rm U}(T)$ at $\kappa \approx \kappa_{ac}$ (a) and
     $\kappa=7.4605$ (b).} \label{fig:ACI_tunnel_evolve}
\end{figure}

In the region of the avoided crossing, the odd-parity state $|\phi_{\gamma}
\rangle$ will be approximately unchanged, while states $|\phi_{\alpha}
\rangle$ and $| \phi_{\beta} \rangle$ become, to lowest order in $\lambda$,
superpositions of their unperturbed, Floquet pendulum, counterparts.
Therefore, in the neighborhood of the avoided crossing, both $|\phi_{\alpha}
\rangle$ and $| \phi_{\beta} \rangle$ will have significant support in the
same region of phase space as $|\phi_{\gamma} \rangle$.  At the exact
position $\kappa_{ac}$ of the avoided crossing, where $|
\phi_{\alpha}\rangle$ and $|\phi_{\beta}\rangle$ are an equal superposition
of the unperturbed states, the initial conditions localized at either $p
\approx 5$ or $-5$ can then be written
\begin{equation}
|\psi_{\pm}(0) \rangle = \frac{1}{2} | \phi_{\alpha} \rangle + \frac{1}{2}
| \phi_{\beta} \rangle \pm \frac{1}{\sqrt{2}} | \phi_{\gamma} \rangle
\,,
\end{equation}
where all three eigenstates are evaluated at $\kappa = \kappa_{ac}$.
Applying the time-evolution matrix to $|\psi_+(0)\rangle$ we find,
after $n$ applications,
\begin{align}
| \psi_+(nT) \rangle &= \hat{\rm U}(nT) | \psi_+(0) \rangle \\ &= {\rm
  e}^{-inT \Omega_{\beta}} \left( \frac{{\rm e}^{-i n T
    \Delta \Omega_{\alpha \beta}}}{2} | \phi_{\alpha} \rangle +
\frac{1}{2} | \phi_{\beta} \rangle + \frac{{\rm e}^{-i n T
    \Delta \Omega_{\gamma \beta}}}{\sqrt{2}} | \phi_{\gamma}
\rangle \right)\,, \nonumber
\end{align}
where $\Delta \Omega_{ij} \equiv \Omega_i - \Omega_j$, here.  If we make the
assumption that $|\Delta \Omega_{\beta \gamma}| = |\Delta \Omega_{\alpha
\gamma}| \equiv \Delta$ at $\kappa=\kappa_{ac}$, we see that
$|\psi_+(\frac{\pi}{\Delta}) \rangle \sim |\psi_-(0)\rangle$. This
three-state process therefore generates a new, larger, tunneling frequency
$\omega_{tun}=\Delta$. The time-evolution of $|\psi_+(0) \rangle$ at
$\kappa\approx\kappa_{ac}$ is shown in Figure \ref{fig:ACI_tunnel_evolve}.a.
Notice that the theoretical tunneling period $n_{tun} =
\frac{\omega}{\Delta} \approx 1.8 \times 10^{5}$ is modulated by a beat
period resulting from the small difference $\Delta \Omega_{\beta \gamma} -
\Delta \Omega_{\gamma \alpha}$. As the position of the avoided crossing is
chosen more precisely, and $\Omega_{\gamma} \rightarrow
\frac{\Omega_{\alpha} + \Omega_{\beta}}{2}$, the period of beating goes to
infinity.

Between these two extremes of regular two-state and three-state
tunneling, at other values of the parameter $\kappa$ along the avoided
crossing, the tunneling takes on a beating behavior due to the two
eigenvalue differences between the odd-parity state and each
even-parity state.  An example of this (the evolution of $|\psi_+(0)
\rangle$ under $\hat{\rm U}(T)$ at $\kappa=7.4605$) is shown in Figure
\ref{fig:ACI_tunnel_evolve}.b.  For this and all values of the
parameter $\kappa$, the evolution of the momentum expectation of
$|\psi_+(0) \rangle$ is well fit by the simple function
\begin{equation}
\langle p \rangle(nT) = A_{\alpha} \cos \left( \Delta \Omega_{\gamma \alpha}
n T \right) + A_{\beta} \cos \left( \Delta \Omega_{\beta \gamma} n T \right)
\,,
\end{equation}
where $A_{\alpha}$ and $A_{\beta}$ can be related to the overlap of
$|\psi_+(0)\rangle$ with $|\phi_{\alpha}\rangle$ and $| \phi_{\beta}
\rangle$, respectively.

The variation of two-state tunneling frequencies in the vicinity of avoided
crossings has been remarked on by many authors
\cite{UNTexasI,UNTexasII,grossmann,israel,tomsovic,ullmo,mouchet_delandeII}
in many different systems, and is often attributed to the influence of
underlying classical chaos.  Classical chaos in a non-perturbative regime
will certainly introduce additional complications to the quantum dynamical
tunneling process which we have not investigated here (notably the
interaction between tunneling through dynamical barriers and free evolution
in a region dominated by chaos \cite{podolskiy}); and avoided crossings will
become larger, more numerous, and may overlap, leading to interaction of
more than three relevant states. However, as we have seen in this section,
the basic mechanism for ``tunneling enhancement'' does not necessarily
require global chaos but only the non-integrability which leads to avoided
crossings. Indeed, for the non-integrability parameter used in this section
($\lambda=5 \times 10^{-2}$), the classical phase space has only small
regions of chaos.

\subsection{Analysis of a Tunneling experiment}

As an application of the tunneling results of the previous section, we
consider the experiment of Steck, Oskay and Raizen \cite{steck_raizen}.  The
Hamiltonian implemented in this atom-optics experiment depends on a single
parameter, $\alpha$, and can be considered a special case of that in Eq.
(\ref{eqn:quant_model_scale}) by setting $\kappa \approx \alpha/2.17$ (up to
a sign difference which can be removed by a $\pi$-translation of the angle
variable), $\lambda = \kappa/2$, and $\omega \approx 6.04$.  It should be
noted that this system is not connected to the Floquet pendulum system since
$\lambda$ is not an independent parameter. Instead, in the limit
$\lambda,\kappa \rightarrow 0$, the free particle Hamiltonian is obtained.

\begin{figure}
     \begin{center}
     \includegraphics{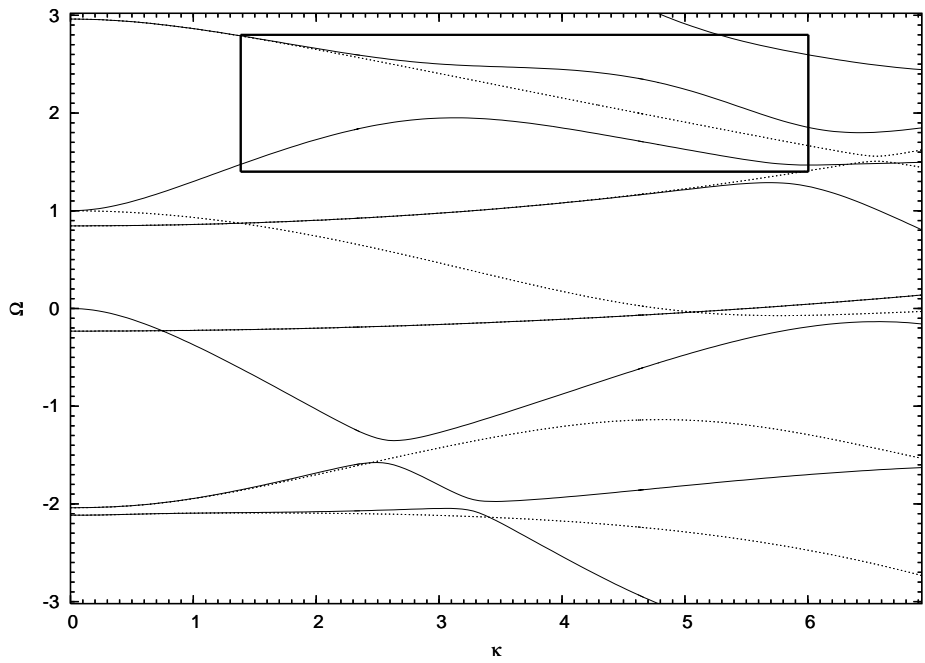}
     \end{center}
     \caption{Floquet eigenvalue curves for the system considered in
\cite{steck_raizen} ($\lambda=\kappa/2$, $\omega \approx 6.04$).}
\label{fig:luter_raizen}
\end{figure}

\begin{figure}
     \begin{center}
     \includegraphics{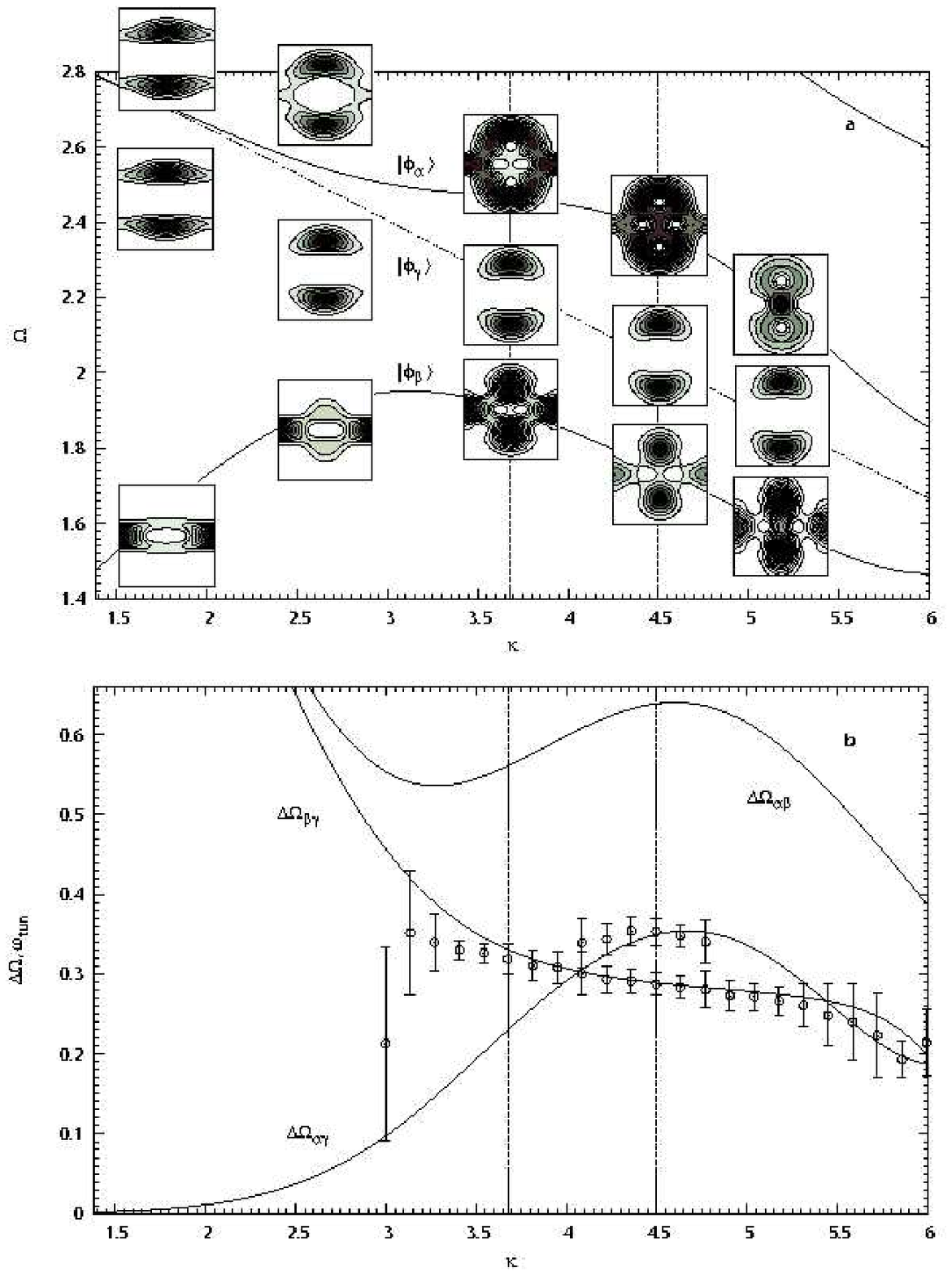}
     \end{center}
     \caption{The boxed region of Figure \ref{fig:luter_raizen} where an
avoided crossing occurs between two even parity states (a). The Husimi
distributions of these and a third, odd-parity, state are overplotted, with
axes of $\theta_0 \in (0,2\pi)$ and $p \in (-6,6)$.  The differences of
these three eigenvalues are shown in (c), with experimental tunneling
frequencies overplotted (circles, reprinted with permission from Steck,
{\em et. al.} \cite{steck_raizen}, Fig. 1).} \label{fig:luter_raizenII}
\end{figure}

The primary result of the experiments detailed in \cite{steck_raizen} is the
observation of dynamical tunneling, between two resonance islands in the
classical phase space (located at $p\approx\pm 3$), which exhibits
oscillation frequencies dependent on the parameter value considered and
independent of the modulation frequency.  In particular, for a range of
parameter values, the observed tunneling oscillations were dominated by two
primary frequencies.  The Floquet eigenvalue curves for the experimental
system are shown in Figure \ref{fig:luter_raizen}, and the experimentally
observed tunneling frequencies are shown in Figure
\ref{fig:luter_raizenII}.b, overplotted on the differences of eigenvalues
between three particular Floquet eigenstates. These three states exhibit
significant localization in the region of the classical resonance islands at
some values of the parameter $\kappa$ \cite{luter_reichl}.  The experimental
frequencies are well predicted by only two of these differences, namely
$\Delta \Omega_{\alpha \gamma}$ and $\Delta \Omega_{\beta \gamma}$.

The oscillations seen in the parameter region $\kappa \in [3,4.75]$ are due
to the existence of an avoided crossing between the eigenvalues of the
even-parity states labelled $|\phi_{\alpha}\rangle$ and $|\phi_{\beta}
\rangle$.  As can be seen in the overplotted Husimi distributions, these two
even-parity states originate at small $\kappa$ values from two disconnected
regions of the phase space, the first residing at $p \approx 0$, and the
other at the positions of the classical resonance islands where the odd
parity state $|\phi_{\gamma}\rangle$ also has its primary support.  In
passing through this avoided crossing, the odd-parity state retains its
original character, while the two even-parity states become mutual
superpositions of their small $\kappa$ character. It is clear that this
avoided crossing is not of the ideal form considered in the perturbative
regime.  The superposing effects of this avoided crossing are extend well
beyond the minimum eigenvalue separation at $\kappa \approx 3.0$ (see Figure
\ref{fig:luter_raizenII}.a), due to the natural eigenvalue curvature of a
state with the low $\kappa$ character of $| \phi_{\beta}\rangle$ (a simpler
example of this is state $(1,0)$ in Figure
\ref{fig:pendulum_to_floquetpend}.b). Just past this minimum spacing, the
eigenvalue of state $|\phi_{\alpha} \rangle$ curves back downward toward
that of $|\phi_{\beta} \rangle$, immediately beginning a second avoided
crossing and thus preserving the composite nature of these two even-parity
states through $\kappa \approx 6$.  As a further complication, in the
parameter region $\kappa \in[5,7]$, states $|\phi_{\alpha} \rangle$ and
$|\phi_{\beta} \rangle$ are joined by two other even-parity states in a
complex and overlapping set of avoided crossings.

\begin{figure}
     \begin{center}
     \includegraphics{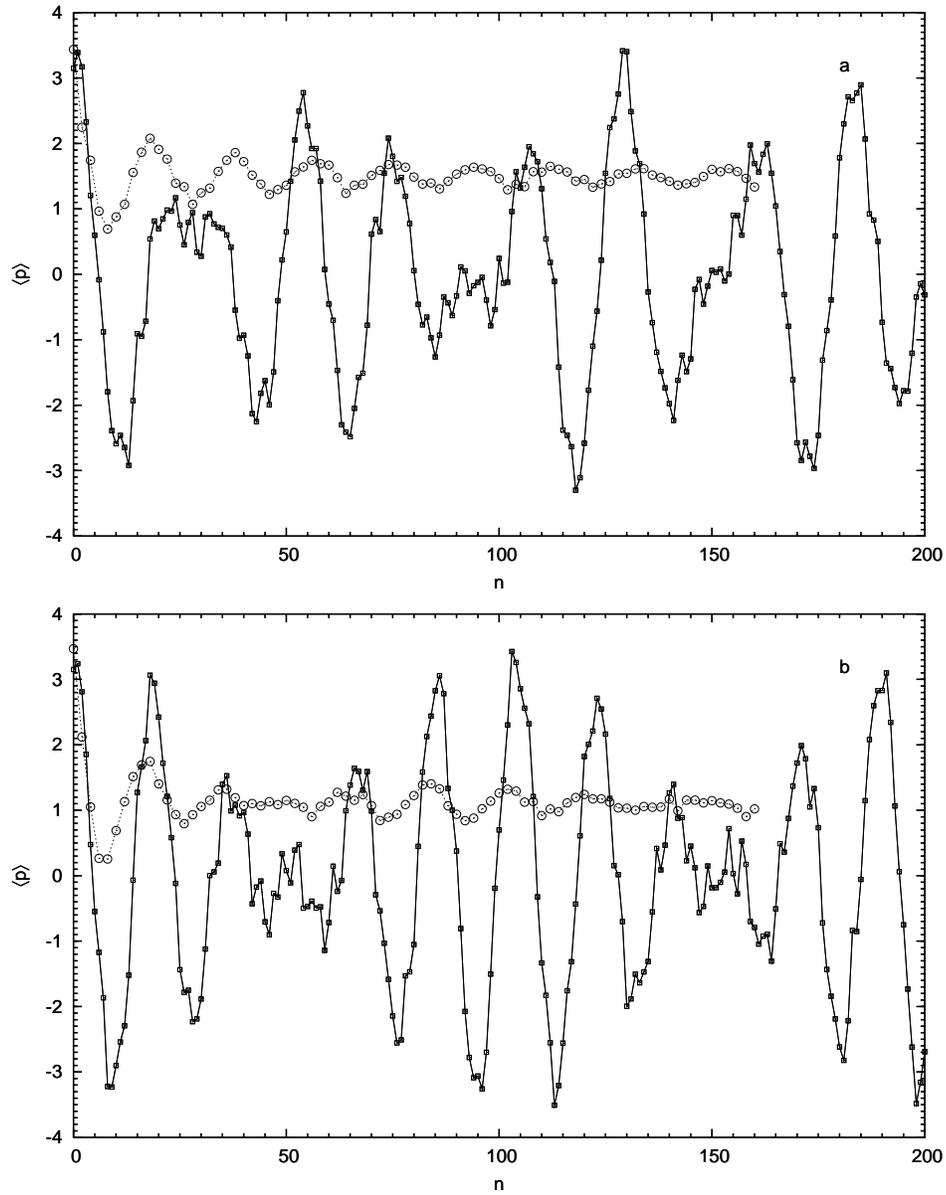}
     \end{center}
     \caption{Numerical evolution (squares) of a positive momentum centered
initial state under the time-evolution operator at $\kappa=3.68$ (a) and
$4.50$ (b), with experimental values overplotted (circles, reprinted with
permission from Steck, {\em et. al.} \cite{steck_raizen}, Fig. 2).}
\label{fig:steck_tunnel}
\end{figure}

Despite these complications, we can identify the dynamical tunneling in the
parameter region $\kappa \in [2.5,4.75]$ to be a three-state process
involving those states shown in Figure \ref{fig:luter_raizenII}.a.  As
predicted by the results of the previous section, the observed tunneling
frequencies involve only differences in Floquet eigenvalue between the odd
parity state and the two even parity states.  A direct comparison of the
oscillations from \cite{steck_raizen} with the numerically calculated
evolution of an initially localized state $|\psi_+(0)\rangle$ under
$\hat{\rm U}(T)$ is shown in Figure \ref{fig:steck_tunnel} for the values
$\kappa = 3.68$ and $4.50$.  Neglecting dissipation and a momentum offset of
the experimental values (due to the fact that not all atoms contributing to
the average are participating in the dynamics), there is good agreement in
the second case.  In the first case, and for all parameter values between
$\kappa=3$ and $4$, the experiment seems to pick up only one of the
underlying frequencies ($\Delta \Omega_{\gamma \beta}$, while the numerics
predict nearly equal contributions from $\Delta \Omega_{\gamma\beta}$ and
$\Delta \Omega_{\alpha \gamma}$.  It was noted in Reference
\cite{luter_reichlII} that the detection of fewer than the predicted number
of frequency contributions to the tunneling behavior was also found in
another experimental system \cite{nist1}.

\begin{figure}
     \begin{center}
     \includegraphics{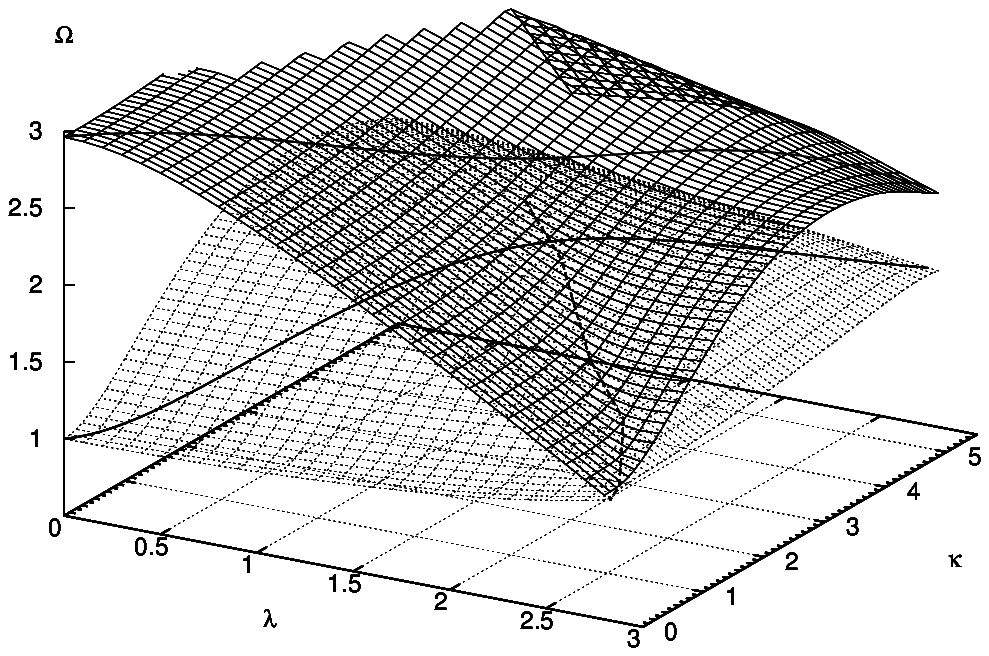}
     \end{center}
     \caption{Eigenvalue surfaces, over $\kappa-\lambda$ space, of the even
parity Floquet eigenstates involved in the avoided crossing of Figure
\ref{fig:luter_raizenII}.a. The eigenvalues on the particular curve
investigated in reference \cite{steck_raizen} are overplotted in bold. A
dashed line traces the minimum separation of the two surfaces from the
diabolical point at $\kappa=0$, $\lambda \approx 2.8$ to the avoided
crossing seen in the experiment.}
\label{fig:luter_diabolical}
\end{figure}

Finally, we would like to consider the origin of the avoided crossing
involved in the dynamical tunneling observed in \cite{steck_raizen}. Figure
\ref{fig:luter_diabolical} shows the avoided crossing of Figure
\ref{fig:luter_raizenII}.a lying on the eigenvalue surfaces
$\Omega_{\alpha}$ and $\Omega_{\beta}$ in $\kappa-\lambda$ space.  One can
see that these two surfaces meet at a diabolical point on the $\kappa=0$
axis, where $\lambda \approx 2.8$.  Numerical analysis shows that, {\em in
the neighborhood of the diabolical point}, the minimum spacing between the
two eigenvalue surfaces is linearly dependent on $\kappa$, with
\begin{equation}
\Delta_{\alpha \beta} = 9.22 \times 10^{-2} \kappa \,.
\end{equation}
This result is in good agreement with a perturbation analysis similar to
that of Appendix A, but with $\kappa$ as the small expansion parameter.  The
predicted minimum spacing from such an analysis yields
\begin{equation}
\Delta_{\alpha \beta,{\rm pt}} = 2 | V^{\prime}_{\alpha \beta} | \kappa \approx 9.2 \times
10^{-2} \kappa\,,
\label{eqn:diabolical_pt}
\end{equation}
where the matrix element must be numerically calculated as the coupling of
the two degenerate Floquet eigenstates of the ($\kappa=0.0$,$\lambda=2.8$)
system through $\hat{V}^{\prime} = \cos \hat{\theta}$, the coefficient of
$\kappa$.  Although the avoided crossing in the experiment (Figure
\ref{fig:luter_raizenII}.a) appears at a parameter value outside the range
of validity of perturbation theory, the observed minimum spacing between
$\Delta \Omega_{\alpha}$ and $\Delta \Omega_{\beta}$ agrees with Eq.
\ref{eqn:diabolical_pt} to within a factor of two.
\section{\label{sec:conclusions}Conclusions}
We have shown that the minimum spacing of avoided crossings in a
near-integrable time-periodic system, and therefore the minimally separated
``ridges'' of the double-cone structures surrounding a diabolical point,
exhibits a power law dependence on the non-integrability parameter, with
integer power.  A modified degenerate perturbation theory has allowed us to
relate the coefficient of this dependence to the direct or indirect coupling
of the two related unperturbed Floquet states through the perturbation
operator, and the integer power to the number of ``photon energies''
$\omega$ connecting their related energy eigenvalues.  Moreover, the
perturbation analysis predicts a qualitatively identical structure for all
avoided crossings which allows us to characterize generically their affect
on dynamical tunneling.  This description was applied to a particular
avoided crossing generating multiple-state tunneling oscillations in an
experimental system, and its connection to a nearby diabolical point was
revealed. It is hoped that these results will provide guidance for the
development of new experimental setups which intend to use avoided crossings
in the realization of multiple state tunneling processes and adiabatic
transitions, or the preparation of Schr\"odinger's cat type superposed
states.

\section{Acknowledgements}

The authors thank the Robert A. Welch Foundation (Grant No. F-1051)
and the Engineering Research Program of the Office of Basic Energy
Sciences at the U.S. Department of Energy (Grant
No. DE-FG03-94ER14465) for support of this work. Author LER thanks the
Office of Naval Research (Grant No.  N00014-03-1-0639) for partial
support of this work. The authors also thank Robert Luter for many
helpful discussions and Daniel Steck for providing us with the
experimental data from reference \cite{steck_raizen}.

\appendix
\section{\label{sec:pert_th}Perturbation Theory for Avoided Crossings}
The Floquet pendulum system $\hat{H}^0_F(\kappa)$ has a two-fold
degeneracy at $\kappa=\kappa_0$ where two eigenstates $| \alpha^0
\rangle \equiv | n_{\alpha}(\kappa_0) , q_{\alpha} \rangle$ and $|
\beta^0 \rangle \equiv | n_{\beta}(\kappa_0) , q_{\beta} \rangle$ have
eigenvalues $\Omega_{\alpha}(\kappa_0) =\Omega_{\beta}(\kappa_0)$. We
will use a modified degenerate perturbation theory to lift this
degeneracy at the $\kappa$-position of the resulting avoided crossing
when $\lambda>0$. In light of the discussion in Section
\ref{sec:pert_th_res}, we write the Floquet Hamiltonian as
\begin{equation}\label{eqn:modified_HF}
{\hat H}_F(\kappa)={\hat H}_F^0+{\delta {\hat
H}_F}\; \Delta\kappa(\lambda)+{\lambda}\;{\hat V},
\end{equation}
where we now write ${\hat H}_F^0 \equiv {\hat H}_F^0(\kappa_0)$. The
operator ${\delta {\hat H}_F}$ is $\frac{\partial
\hat{H}_F}{\partial \kappa}=\cos\hat{\theta}$ with the
simplifying assumption that this operator is diagonal in the
unperturbed basis. This is equivalent to the assumption that the
unperturbed eigenvalue curves are linear in the small region of
$\kappa$ under consideration. We expand $\Delta\kappa(\lambda)$,
setting $\Delta\kappa (0) = 0$, so that
\begin{equation}\label{eqn:dkappa_expand}
\Delta\kappa(\lambda)=\Delta\kappa^{(1)}{\lambda}+\Delta\kappa^{(2)}{\lambda}^2+
\cdots
\end{equation}
We expand the near-degenerate eigenstates and eigenvalues in powers of
$\lambda$ about their ($\kappa=\kappa_0,\lambda=0$) values:
\begin{equation}\label{eqn:pert_expand}
\begin{split}
|\phi{\rangle}&=C_{\alpha}|\alpha^0{\rangle}+C_{\beta}|\beta^0{\rangle}
+{\lambda}|\phi^1{\rangle} +{\lambda}^2|\phi^2{\rangle}+ \cdots \\
{\Omega}&={\Omega}^{(0)}+{\lambda}{\Omega}^{(1)} +{\lambda}^2{\Omega}^{(2)}+ \cdots
\end{split}
\end{equation}
where $\hat{H}^0_F | \alpha^0 \rangle = \Omega^{(0)} | \alpha^0
\rangle$, $\hat{H}^0_F | \beta^0 \rangle = \Omega^{(0)} | \beta^0
\rangle$ and the zeroth-order eigenstates have been assumed to be a
superposition of the two degenerate unperturbed states.  As is usual
in degenerate perturbation theory, the lowest-order near-degenerate
eigenstates and the corrections to their eigenvalues will be the
eigenvectors and eigenvalues of a $2 \times 2$ matrix in the basis of
the two degenerate unperturbed states.  At this order, we will make the
distinct assignments $|\phi \rangle \rightarrow |\phi_{\pm} \rangle$
(corresponding to the solutions of the quadratic characteristic
equation). The final expressions for $C^{\pm}_{\alpha}$,
$C^{\pm}_{\beta}$, $|\phi^i_{\pm} \rangle$, and $\Omega^{(i>0)}_{\pm}$
will depend on $\kappa$ through the particular choice of the $\Delta
\kappa^{(i)}$'s. The Floquet eigenvalue equation (\ref{eqn:floq_ham})
for the near-degenerate state now takes the form
\begin{multline} \label{eqn:degenpert_full}
\left[ (\hat{H}_F^0 -
{\Omega}^{(0)})~\left(C_{\alpha}|\alpha^0{\rangle}
+C_{\beta}|\beta^0{\rangle}\right)\right] \\
+ \lambda \left[ (\hat{H}_F^0 -
{\Omega}^{(0)})|{\phi^1}{\rangle}+\left(\Delta\kappa^{(1)} {\delta {\hat
H}_F}+ \hat{V} -{\Omega}^{(1)}\right)~
\left(C_{\alpha}|\alpha^0{\rangle}+C_{\beta}|\beta^0{\rangle}\right)\right]
\\
+ \lambda^2 \left[(\hat{H}_F^0 -{\Omega}^{(0)}) |{\phi^2}{\rangle} +
(\Delta\kappa^{(1)} {\delta {\hat H}_F} + \hat{V} -{\Omega}^{(1)})
|{\phi^1}{\rangle} \right.\\
\left. + (\Delta\kappa^{(2)} {\delta {\hat H}_F} -
{\Omega}^{(2)})~(C_{\alpha}|\alpha^0{\rangle}
+C_{\beta}|\beta^0{\rangle}) \right] + \cdots = 0 \,.
\end{multline}
\subsection{First-order results}

At first order in $\lambda$ we have the eigenvalue equation
\begin{equation}
      (\hat{H}_F^0 -
{\Omega}^{(0)})|{\phi^1}{\rangle}+(\Delta\kappa^{(1)}
{\delta {\hat H}_F}+
\hat{V} -{\Omega}^{(1)}) (C_{\alpha}|\alpha^0{\rangle}
+C_{\beta}|\beta^0{\rangle}) = 0 \,.
\end{equation}
If we now act on this equation with
${\langle}\alpha^0|$ and then with ${\langle}\beta^0|$, we obtain
\begin{equation}\label{eqn:firstorder-1}
     \left( \begin{array}{cc}
     \Delta\kappa^{(1)} {\delta}E_{\alpha} + V_{\alpha \alpha} -
{\Omega}^{(1)} & V_{\alpha \beta} \\
     V_{\beta \alpha} & \Delta\kappa^{(1)}{\delta}E_{\beta} +
V_{\beta \beta} -
{\Omega}^{(1)}
\\
     \end{array} \right)
     \left( \begin{array}{c}
     C_{\alpha} \\
     C_{\beta}  \\
     \end{array} \right) = {\bf 0} \,.
\end{equation}
where $V_{ij}$ are the matrix elements of the perturbation in the
basis of the degenerate unperturbed eigenstates,
\begin{equation}
\delta E_i \equiv \langle \langle n_i,q_i | \delta
\hat{H}_F | n_i,q_i \rangle \rangle = \langle n_i | \cos
\hat{\theta}| n_i \rangle \,,
\end{equation}
and we have used the identity
\begin{equation}
1 = \sum_{\gamma} |\gamma^0 \rangle \langle \gamma^0 | \equiv
\sum_{\left(n_{\gamma},q_{\gamma}\right)} |n_{\gamma},q_{\gamma}\rangle \langle
n_{\gamma},q_{\gamma} | = \sum_{n,q = -\infty}^{\infty} | n,q \rangle
\langle n,q | \,.
\end{equation}
At this point, we must consider two possible cases: $V_{\alpha \beta} \ne 0$
and $V_{\alpha \beta} = 0$. In the first case, a nontrivial solution for the
$C_i$ requires first-order corrections to the unperturbed Floquet eigenvalues
\begin{multline}
\label{eqn:omega1pm-1}
     {\Omega}^{(1)}_{\pm} = \frac{1}{2} \left[ V_{\alpha \alpha} +
V_{\beta \beta} + \Delta \kappa^{(1)}
     ({\delta}E_{\alpha} +{\delta}E_{\beta})\right] \\
     \pm \frac{1}{2} \left\{ \left[(V_{\alpha \alpha} -
     V_{\beta \beta}) + \Delta\kappa^{(1)} ({\delta}E_{\alpha} -
{\delta}E_{\beta})
\right]^2 + 4 |V_{\alpha \beta}|^2
     \right\}^{\frac{1}{2}}\,.
\end{multline}
We can then write the spacing between the two Floquet eigenvalues
\begin{equation}
     \Delta{\Omega} {\equiv}
     {\Omega}^{(1)}_+ -
     {\Omega}^{(1)}_- = \lambda \left\{ \left[(V_{\alpha \alpha} -
     V_{\beta \beta}) + \Delta\kappa^{(1)} ({\delta}E_{\alpha}
-{\delta}E_{\beta})
\right]^2 + 4 |V_{\alpha \beta}|^2
     \right\}^{\frac{1}{2}} + O(\lambda^2)\,.
\label{space-1}
\end{equation}
In order to determine the particular value of $\Delta\kappa^{(1)}$ at which
the minimum spacing occurs, we solve the extremal equation
\begin{equation}
     \left(\frac{\partial \Delta\Omega}{\partial \Delta\kappa}\right)_\lambda
     = \frac{1}{\lambda} \left(\frac{\partial \Delta\Omega}{\partial
     \Delta\kappa^{(1)}}\right)_{\lambda,\Delta\kappa^{(m)}}= 0 \quad (m \ne 1)
     \,,
\end{equation}
finding that
\begin{equation}
     \Delta\kappa^{(1)}_{ac} = \frac{V_{\beta \beta} -
V_{\alpha \alpha}}{{\delta}E_{\alpha}-{\delta}E_{\beta}}
     \,.
\end{equation}
If we substitute this into Eq. (\ref{eqn:omega1pm-1}), we obtain
first order corrections to the eigenvalues of
\begin{equation}\label{eqn:omega1pm-2}
\Omega^{(1)}_{\pm} = \frac{V_{\beta \beta} \delta E_{\alpha} -
V_{\alpha \alpha} \delta E_{\beta}}{\delta E_{\alpha} -\delta
E_{\beta}} \pm | V_{\alpha \beta}|\,.
\end{equation}
Therefore, we find the minimum splitting to be
\begin{equation}
     \Delta_{\alpha \beta} = 2 \lambda |V_{\alpha \beta}| + O(\lambda^2)\,.
\end{equation}

In the second case ($V_{\alpha \beta} = 0$), we find that the nearest approach of
the two Floquet eigenvalues is in fact a {\it crossing} (to first order in
$\lambda$).  In this case we define $\Delta\kappa^{(1)}_{ac}$ to be the
offset of this crossing, i.e.
\begin{equation}
    {\Delta}{\Omega}= O(\lambda^2)~~~{\rm at}~~~
\Delta\kappa^{(1)}_{ac} \equiv \frac{V_{\beta \beta} -
V_{\alpha \alpha}}{{\delta}E_{\alpha}-{\delta}E_{\beta}}.
\end{equation}
The coefficients $C_i$ remain undetermined until the degeneracy
is broken.
\subsection{Second-order results ($V_{\alpha \beta} = 0$)}

For the case $V_{\alpha \beta}=0$, the zeroth order states must be
determined from the second order equation, which takes the form
\begin{eqnarray}
     (\hat{H}_F^0  -{\Omega}^{(0)}) |{\phi^2}{\rangle}
     + (\Delta\kappa^{(1)} {\delta {\hat H}_F} + \hat{V}
-{\Omega}^{(1)})
|{\phi^1}{\rangle} \nonumber\\
     + (\Delta\kappa^{(2)} {\delta {\hat H}_F}
  -
{\Omega}^{(2)})~(C_{\alpha}|\alpha^0{\rangle}
+C_{\beta}|\beta^0{\rangle}) = 0\,,
     \label{eqn:second_ord_pert}
\end{eqnarray}
Following the same procedure as in first order, we obtain
\begin{equation}
     \left( \begin{array}{cc}
     \Delta\kappa^{(2)} {\delta}E_{\alpha} +{v}_{\alpha \alpha}
- {\Omega}^{(2)} &
{v}_{\alpha \beta} \\
     {v}_{\beta \alpha} & \Delta\kappa^{(2)} {\delta}E_{\beta} +
{v}_{\beta \beta} -
{\Omega}^{(2)}
\\
     \end{array} \right)
     \left( \begin{array}{c}
     C_{\alpha} \\
     C_{\beta} \\
     \end{array} \right) = {\bf 0} \,,
\end{equation}
where we have used the first-order result
\begin{equation} \label{eqn:gamma0_phi1}
\langle \gamma^0 | \phi^1 \rangle = \frac{V_{\gamma \alpha} C_{\alpha} +
V_{\gamma \beta} C_{\beta}}{\Omega^{(0)} - \Omega^{(0)}_{\gamma}} \quad
\left(\gamma \not\in \{\alpha, \beta\}\right)
\end{equation}
with $\Omega^{(0)}_{\gamma} = \langle n_{\gamma}, q_{\gamma} | \hat{H}_F^0 |
n_{\gamma}, q_{\gamma} \rangle$, and where we have defined
\begin{equation}
     {v}_{ij} \equiv \sum_{\gamma \not \in \{\alpha,\beta\}} \frac{V_{i
\gamma}V_{\gamma j}}{\Omega^{(0)} - {\Omega}^{(0)}_{\gamma}} \quad \left(
i,j \in \{\alpha,\beta\}\right)\,.
\end{equation}
Again we must consider two cases: ${v}_{\alpha \beta} \ne 0$ and
${v}_{\alpha \beta} = 0$. In the first case, our procedure for
determining $\Delta\kappa^{(2)}_{ac}$ is identical to that of
first-order and we obtain
\begin{equation}
     \Delta\kappa^{(2)}_{ac} = \frac{{v}_{\beta\beta} -
{v}_{\alpha \alpha}}{{\delta}E_{\alpha}
     -{\delta}E_{\beta}} \quad {\rm and} \quad \Delta_{\alpha \beta}
     = 2 \lambda^2 |{v}_{\alpha \beta}| + O(\lambda^3)\,.
     \label{eqn:lambda-sqd_offset}
\end{equation}
In the case that $v_{\alpha \beta}=0$, we find that ${\Delta}{\Omega}=
O(\lambda^3)$ at an offset of
\begin{equation}
     \Delta\kappa^{(2)}_{ac} = \frac{{v}_{\beta\beta} - {v}_{\alpha
     \alpha}}{{\delta}E_{\alpha} -{\delta}E_{\beta}} \,.
\end{equation}
\subsection{Third-order results ($V_{\alpha \beta}=0$ and
$v_{\alpha \beta}=0$)}

If the conditions $V_{\alpha \beta}=0$ and $v_{\alpha \beta}=0$ are
satisfied, we can attempt to lift the degeneracy at order $\lambda^3$.
We obtain the following results
\begin{equation}
     \left( \begin{array}{cc}
         \Delta\kappa^{(3)} {\delta}E_{\alpha}
     +\mathfrak{v}_{\alpha \alpha} + \bar{\mathfrak{v}}_{\alpha} -
     {\Omega}^{(3)} & \mathfrak{v}_{\alpha \beta} \\ \mathfrak{v}_{\beta
     \alpha} & \Delta\kappa^{(3)} {\delta}E_{\beta} +
     \mathfrak{v}_{\beta \beta} + \bar{\mathfrak{v}}_{\beta} -
     {\Omega}^{(3)} \\
     \end{array} \right)
     \left( \begin{array}{c}
     C_{\alpha} \\
     C_{\beta} \\
     \end{array} \right) = {\bf 0} \,,
\end{equation}
where we have defined
\begin{equation}
\mathfrak{v}_{ij} \equiv \sum_{\gamma,\sigma \not \in \{\alpha,\beta\}}
\frac{V_{i \gamma}V_{\gamma \sigma} V_{\sigma j}}{\left(\Omega^{(0)} -
{\Omega}^{(0)}_{\gamma}\right)\left(\Omega^{(0)} -
\Omega^{(0)}_{\sigma}\right)} \quad \left( i,j \in \{\alpha,\beta\}\right)\,,
\end{equation}
and
\begin{equation}
\bar{\mathfrak{v}}_i \equiv \sum_{\gamma \not \in \{\alpha,\beta\}}
\frac{\left( \Delta\kappa^{(1)} \delta E_{\gamma} - \Omega^{(1)}\right)
V_{i \gamma}V_{\gamma i}}{\left(\Omega^{(0)} -
{\Omega}^{(0)}_{\gamma}\right)^2} \quad \left( i \in
\{\alpha,\beta\}\right)\,,
\end{equation}
and where we have used an expression for $\langle \gamma^0 | \phi^2 \rangle$
(analogous to Equation \ref{eqn:gamma0_phi1}) determined from the second-order
equation.

These equations are nearly of the same form as those at first and second
order.  By the same procedure we determine
\begin{equation}
     \Delta\kappa^{(3)}_{ac} = \frac{\mathfrak{v}_{\beta\beta} +
\bar{\mathfrak{v}}_{\beta} -
\mathfrak{v}_{\alpha \alpha} - \bar{\mathfrak{v}}_{\alpha}}{{\delta}E_{\alpha}
     -{\delta}E_{\beta}} \quad {\rm and} \quad \Delta_{\alpha \beta}
     = 2 \lambda^3 |\mathfrak{v}_{\alpha \beta}| + O(\lambda^4)\,,
     \label{eqn:lambda-cubed_offset}
\end{equation}
for the case that $\mathfrak{v}_{\alpha \beta} \ne 0$ and $\Delta_{\alpha
\beta} = O(\lambda^4)$ when $\mathfrak{v}_{\alpha \beta} = 0$.

\pagebreak

\end{document}